%% file: paper_Lin_Cub_Soc_V3_9_04_2018.tex
\documentclass[twocolumn,english,superscriptaddress,aps,prl]{revtex4-1}
\usepackage{lmodern}
\usepackage[T1]{fontenc}
\usepackage[latin9]{inputenc}
\usepackage{color}
\usepackage{babel}
\usepackage{mathrsfs}
\usepackage{amsmath}
\usepackage{amssymb}
\usepackage{graphicx}
\usepackage[export]{adjustbox}
\usepackage{esint}
\usepackage[caption=false]{subfig}
\usepackage[normalem]{ulem}
\usepackage{bm} 
\usepackage[unicode=true,pdfusetitle,
 bookmarks=false,
 breaklinks=false,pdfborder={0 0 1},backref=false,colorlinks=true]{hyperref}

\hypersetup{
 colorlinks,linkcolor=blue,citecolor=blue,urlcolor=blue}
\makeatletter

\newcommand{\commentout}[1]{}

\makeatletter
\newlength{\sfp@hseplen}\newlength{\sfp@vseplen}
\define@cmdkey{subfigpos}[sfp@]{pos}[ul]{}
\define@cmdkey{subfigpos}[sfp@]{font}[\small]{}
\define@cmdkey{subfigpos}[sfp@]{vsep}[2.2\baselineskip]{\setlength{\sfp@vseplen}{\sfp@vsep}}
\define@cmdkey{subfigpos}[sfp@]{hsep}[12pt]{\setlength{\sfp@hseplen}{\sfp@hsep}}

\newcommand{\subfigimg}[3][,]{%
	\setkeys{Gin,subfigpos}{pos,font,vsep,hsep,#1}
	\setbox1=\hbox{\includegraphics{#3}}
	\ifnum\pdfstrcmp{\sfp@pos}{ul}=0
	\leavevmode\rlap{\usebox1}
	\rlap{\hspace*{\sfp@hsep}\raisebox{\dimexpr\ht1-\sfp@vsep}{\sfp@font{#2}}}
	\phantom{\usebox1}
	\else\ifnum\pdfstrcmp{\sfp@pos}{ur}=0
	\leavevmode\usebox1
	\llap{\raisebox{\dimexpr\ht1-\sfp@vsep}{\sfp@font{#2}}\hspace*{\sfp@hsep}}
	\else\ifnum\pdfstrcmp{\sfp@pos}{lr}=0
	\leavevmode\usebox1
	\llap{\raisebox{\sfp@vsep}{\sfp@font{#2}}\hspace*{\sfp@hsep}}
	\else
	\leavevmode\rlap{\usebox1}
	\rlap{\hspace*{\sfp@hseplen}\raisebox{\sfp@vsep}{\sfp@font{#2}}}
	\phantom{\usebox1}
	\fi\fi\fi
}
\makeatother

\begin{document}

\title{Theory of the inverse spin galvanic effect in quantum wells}
\author{Amin Maleki Sheikhabadi}
\affiliation{Dipartimento di Matematica e Fisica, Università Roma Tre, 00146 Rome, Italy }
\affiliation{Department of Physics, Kent State University, 44242, Kent, Ohio, USA}
\author{Iryna Miatka}
\affiliation{Dipartimento di Matematica e Fisica, Università Roma Tre, 00146 Rome, Italy }
\author{E. Ya. Sherman}
\affiliation{Department of Physical Chemistry, The University of the Basque Country UPV/EHU, 48940, Leioa, Spain}
\affiliation{IKERBASQUE Basque Foundation for Science, Bilbao, Spain}
\author{Roberto Raimondi}
\affiliation{Dipartimento di Matematica e Fisica, Università Roma Tre, 00146 Rome, Italy }

\begin{abstract}
The understanding of the fundamentals of spin and charge densities and currents interconversion by spin-orbit coupling
can enable efficient applications beyond the possibilities offered by conventional electronics. 
For this purpose we consider various forms of the frequency-dependent inverse 
spin galvanic effect (ISGE) in semiconductor quantum wells and epilayers taking into account
the cubic in the electron momentum spin-orbit coupling in the Rashba and Dresselhaus forms, concentrating on the 
current-induced spin polarization (CISP). We find that including the cubic terms qualitatively explains recent findings
of  the CISP in InGaAs epilayers being  the strongest if the internal spin-orbit coupling field is the smallest and 
vice versa \cite{norman2014,luengo2017}, 
in contrast to the common understanding. Our results provide a promising framework for the control 
of spin transport in future spintronics devices.

\end{abstract}

\date{\today}    

\maketitle


\section{Introduction}

The spin galvanic effect (SGE) and its Onsager reciprocal effect are
currently the focus of an intense investigation in a large variety of
physical systems including metals, semiconductors, van der Waals
heterostructures and topological insulators~\cite{ando_shiraishi2017,fert2016,pesin_macdonald2012,fert_pumping2016,ivchenko2017}. The effect allows the ``spin-to-charge interconversion'',  where a non-equilibrium spin polarization
yields an electrical current (SGE) and, conversely, an applied electrical
current is able to orient the electron spin producing the ISGE. In the latter case one speaks also of CISP. In the literature different names
refer to the same effect, often depending on the context where the
phenomenon is being investigated. A discussion about the nomenclature can be
found in Ref.~\cite{Ganichev2016}. On symmetry grounds the SGE arises when, due to restricted symmetry conditions as in
gyrotropic media~\cite{Ivchenko1978,kokurin2017}, specific components of polar and axial vectors transform
according to the same representation. On a microscopic level, instead, the lack of
inversion symmetry lifts the spin degeneracy leading to a momentum dependent
spin splitting, which acts as an internal effective magnetic field. As a
consequence Bloch electron states have their spin quantization axis dependent on the momentum direction. This aspect gives rise to a well defined spin
texture around the Fermi surface, which can be experimentally
measured, for instance, by the spin-pumping technique~\cite{Sanchez13,chen2016,yoshichika2016,fert_pumping2016} 
and by pump-probe techniques as in semiconducting
epilayers~\cite{norman2014,luengo2017,Ivchenko_chapterBook2017}. After pumping polarized radiation
into the electron system, one can observe a degree of precession of the
induced spin polarization in the internal magnetic field. An essential
ingredient is the external electric field, which unbalances the occupation of
momentum states, yielding a net internal field. In a semiconducting epilayer
the spin-orbit coupling (SOC) acts via two microscopic mechanisms. At the
bulk level, the lack of inversion symmetry of the lattice as in GaAs
heterostructures is responsible for the Dresselhaus term~\cite{Dresselhaus55}, which depends on the 
third power of the electron momentum. However, when
the electron system is confined in one direction, say along the \textit{z} axis, the
Dresselhaus spin-orbit coupling (DSOC) becomes linear in momentum. On the other hand the lack of
inversion symmetry with respect to the growth direction of the epilayer yields the Rashba term~\cite{Rashba84}, which is linear in the momentum and
in the spin operators. A combination of the linear DSOC and the Rashba spin-orbit coupling (RSOC) 
produces a characteristic spin texture, where maximum and
minimum values of the internal field align along the [1,1] and [1,-1]
crystallographic axes, depending on the strength of the two types of SOC.
The pump-probe technique used in Refs.~\cite{norman2014,luengo2017}
is capable of reconstructing the texture of the internal magnetic field by
varying the direction of the applied electric field, thus allowing the
measurement of the DSOC and RSOC.

Theoretical investigations of both linear DSOC and RSOC in a two-dimensional electron 
gas (2DEG)~\cite{Aronov89,Edelstein1990} concluded that the
induced spin polarization is proportional to the internal magnetic field
and, hence, the former aligns with the latter. However, the experimental results of
Refs. \cite{norman2014,luengo2017} showed the opposite behavior: the
maximum spin polarization occurs in correspondence of the minimum value of
the internal field and vice versa. Based on the model developed in 
Refs. \cite{paper2_physB,malekiinverse}, a possible explanation has been 
proposed in Ref. \cite{luengo2017}, by allowing SOC also from
random impurities (see also Refs. \cite{glazov2010,dugaev2010}). 
The latter have a two-fold effect. On the one hand, they
introduce a second channel for spin relaxation, referred to as the
Elliott-Yafet mechanism, in addition to the Dyakonov-Perel (DP) one associated
with the linear RSOC and DSOC. On the other hand, as found in Refs. 
\cite{paper2_physB,malekiinverse}, the interplay of linear RSOC and
DSOC with the impurity SOC yields a negative SGE, which tends to decrease the induced spin polarization described in Refs. \cite{Aronov89,Edelstein1990}.

The aim of this paper is to study theoretically the experimentally  relevant regimes of the ISGE in semiconductor 
structures, including the dependence of the spin polarization on the frequency of the driving electric
field. The understanding of the frequency-dependent response allows one to set the limits on the timescale
of the spin control by the electric field. 
We extend our studies beyond the conventional diffusive regime, that is  to the case when the spin precession rate 
due to the spin-orbit coupling is of the order of the impurity-determined scattering rate, as
can be achieved in modern high-mobility structures (see Ref. \cite{griesbeck2009} as an example). To correspond to the 
experimental realizations, in addition to the linear-in-the electron  
momentum spin-orbit coupling, we include the cubic terms in the Hamiltonian. These SOC terms are 
important for weak antilocalization in quantum wells~\cite{cubicSOC}
and for the persistent spin helix dynamics \cite{luffe2011,detwiller2017}.
We demonstrate that the unusual experimental results of Refs.~\cite{norman2014,luengo2017}, can be explained by
taking into account this cubic SOC.  Indeed, the steady current-induced 
spin density is controlled by the balancing of the spin-generation and
spin-relaxation torques. In the absence of cubic SOC, the linear RSOC and the
DSOC contribute to both torques and, as a result, one obtains the alignment
of the spin polarization along the internal SOC field. Our results demonstrate in detail that the 
cubic SOC, by itself, can only affect the spin-relaxation torques without inducing a spin-generation torque. 
When both linear and cubic SOCs are present, the generation- and relaxation- torques are affected
differently and the spin polarization is no longer bound to align along the
internal field, corresponding to the results of Refs.~\cite{norman2014,luengo2017}.

The layout of the paper is as follows. In Section II we introduce the
formalism based on the Eilenberger equation for the quasiclassical Green
function. In Section III we apply this formalism to the evaluation of the ISGE in the
case of linear RSOC and DSOC  and study its frequency dependence within and beyond the diffusive regime. 
Section IV demonstrates the absence of the ISGE when
only the cubic SOC is present. In Section V we consider the interplay between linear and cubic RSOC and in Section 
VI we analyze the ISGE in a system where
both linear RSOC and DSOC are present together with cubic DSOC. The Section VII presents the conclusions
and relation to the experiment. Some details of the calculations are provided in the Appendices.

\section{The Eilenberger equation}

We consider electrons confined in a two-dimensional $(xy)$ plane  subject to
impurity scattering and in the presence of SOC. The Hamiltonian of the model
in the presence of a generic intrinsic SOC has the form 
\begin{equation}  \label{hamiltonian}
H=\frac{p^2}{2m}+\mathbf{b}\cdot{\bm\sigma}+V(\mathbf{r}),
\end{equation}
where $V(\mathbf{r})$ and $\mathbf{p}=(p_x,p_y)$ represent the impurity
potential and the vector of the momentum, respectively. The random potential has zero average and $\langle V({\bf r})V({\bf r'})\rangle=\delta({\bf r}-{\bf r'})n_iv_0^2$, with $v_0$ being the single-impurity scattering amplitude and $n_i$ being the  impurity concentration.
In the following, we choose units such that $\hbar= 1$ for the sake of simplicity. 
The vector $\mathbf{b}$ can be defined as the effective magnetic
field due to the Rashba-Dresselhaus SOC. In Ref. [\onlinecite{quasiclassical}%
], the Eilenberger equation for the quasiclassical Green function was
derived in the presence of a SOC of the type shown in the Hamiltonian (\ref%
{hamiltonian}). To present a consistent analysis, we first recall the key steps of the derivation. The starting point is
left-right subtracted Dyson equation for the Keldysh Green function $\check{G}
$, which has the form~[\onlinecite{rammer1986}]
\begin{equation}  \label{boltzamn}
\partial_t \check{G} +\frac{1}{2} \Big\{\frac{\mathbf{p}}{m}+\frac{\partial}{%
\partial \mathbf{p}}(\mathbf{b}\cdot {\bm\sigma}),\frac{\partial}{%
\partial \mathbf{x}}\check{G}\Big\}+i [\mathbf{b}\cdot {\bm\sigma},%
\check{G}]=-i[\check{\Sigma}, \check{G}],
\end{equation}
where the self-energy $\check \Sigma$ includes disorder effects and the curly brackets denote the anticommutator. 
In the Wigner coordinates, the Green function is described as $\check{G}=\check{G}(\mathbf{%
p},{\epsilon}, \mathbf{x}, t)$, where $\mathbf{p}$ and ${\epsilon}$ are the
Fourier transform of the relative coordinates $\mathbf{x}_{1}-%
\mathbf{x}_{2}$, $t_{1}-t_{2}$ and $\mathbf{x}=(\mathbf{x}_1+\mathbf{x}_2)/2$, $%
t=(t_1+t_2)/2$ are coordinates of the center of mass. Whenever it is not
strictly necessary, we drop the explicit dependence $\check{G}(\mathbf{p},{\epsilon}%
, \mathbf{x}, t)$ for simplicity's sake. The quasiclassical Green function
is defined as 
\begin{equation}  \label{quassig}
\check{g}=\frac{i}{\pi} \int d \xi \check{G},
\end{equation}
where $\xi=p^2/2m-\mu$ is the energy measured with respect to the chemical
potential $\mu$ in the absence of SOC. For the Green function, following~\cite{quasiclassical} we make the ansatz 
\begin{equation}  \label{green}
\check{G}= 
\begin{bmatrix}
G^R & G^K \\ 
0 & G^A%
\end{bmatrix}%
=\frac{1}{2}\left\{ 
\begin{bmatrix}
G^R_0 & 0 \\ 
0 & -G^A_0%
\end{bmatrix}%
, 
\begin{bmatrix}
\tilde{g}^R & \tilde{g}^K \\ 
0 & \tilde{g}^A%
\end{bmatrix}%
\right\},
\end{equation}
with $G^R_0$ and $G^A_0$ being, respectively, the retarded and advanced Green functions in the absence
of external perturbations 
\begin{equation}
G^{R(A)}_0=\frac{1}{(\epsilon-\xi)\sigma^0 - \mathbf{b} \cdot {\bm\sigma}%
-\Sigma^{R(A)}}
\end{equation}
with the self-energy $\Sigma^{R(A)}$ (derived later) due to the
impurity potential and $\sigma^0$ the identity matrix. According to the ansatz (\ref{green}), in the equilibrium one obtains
\begin{eqnarray}
{\check{\tilde g}}=%
\begin{bmatrix}
1 & 2 \mathrm{tanh} (\epsilon/2T) \\ 
0 & -1%
\end{bmatrix}\otimes\sigma^0%
.  \label{gk}
\end{eqnarray}
Since the main contribution to the $\xi$-integral comes from the domain
$|\xi|\ll\mu$, it is sufficient to expand $\mathbf{b}$ around the small
values of $\xi$. In the limit of $|\mathbf{b}|$ small compared to the Fermi
energy, we have 
\begin{eqnarray}
b\equiv |\mathbf{b}| &\approx& b_0 +\xi \frac{\partial b_0}{\partial \xi}, \\
|p_{\pm}| &\approx& p_F \mp \frac{|b_0|}{v_F},
\end{eqnarray}
where the subscript ``$0$'' denotes the  values taken at the Fermi surface and
the $p_{\pm}$ refers to the Fermi momentum in the $\pm$-band. It is useful
to introduce the projection operators for the two spin subbands as 
\begin{equation}
\mathcal{P}_{\pm}=\frac{1}{2}\left( \sigma^0\pm \mathbf{ b}_0\cdot {%
{{\bm\sigma}}}\right),\qquad \mathbf{b}_0=\mathbf{b}/b.
\label{projectors}
\end{equation}
As a result, the semiclassical Green function $\check g$, defined in Eq.~(\ref{quassig}), can be written as 
\begin{eqnarray}
\check{g}&=&\sum_{\nu=\pm} (1-\nu \partial_{\xi} b_0) \frac{1}{2}\Big\{%
\mathcal{P}_{\nu} ,{\check {\tilde g}}\Big\}\equiv\sum_{\nu=\pm} (1-\nu
\partial_{\xi} b_0) {\check {\tilde g}}_{\nu}  \notag \\
&=&\frac{1}{2}\{\sigma^0-\partial_{\xi}\mathbf{b}_{0}\cdot{\bm\sigma},\check{%
\tilde{g}}\},  \label{shift82}
\end{eqnarray}
from which we find 
\begin{eqnarray}
\check{\tilde{g}}&=&\check{g}+\frac{1}{2}\{\partial_{\xi}\mathbf{b}_{0}\cdot%
{\bm\sigma}, \check{g}\} ,  \label{shift81}
\end{eqnarray}
{where $\partial_\xi$ is the partial derivative taken with respect to $\xi$.}
By means of (\ref{shift82}), one can show that
\begin{equation}  \label{g+-}
\check{g}_{\nu}=\frac{1}{2}\left\{\mathcal{P}_{\nu},\check{g}\right\},\quad 
\check{g}=\sum_{\nu =\pm} \check{g}_{\nu},
\end{equation}
and, moreover, for any function of momentum one obtains: 
\begin{equation}  \label{genericquasig}
\frac{i}{\pi}\int d\xi f(p) \check{G} =\sum_{\nu=\pm}
f(p_{\nu}) {\check g}_{\nu}.
\end{equation}
Eqs.~(\ref{quassig}) and (\ref{genericquasig}), by integrating over the
energy $\xi$ and retaining terms up to the first order in $|\mathbf{b}%
|/\epsilon_F$, allow to derive the Eilenberger equation in the form~\cite{quasiclassical,halperin2006,raimondi_eilenberger2003} 
\begin{eqnarray}
\hspace{-0.4cm}&&\sum_{\nu=\pm}\left[\partial_t \check{g}_{\nu} +\frac{1}{2} \left\{ \left( \frac{{\bf p}_{\nu}%
}{m}+\frac{\partial}{\partial \mathbf{p}}(\mathbf{b}\cdot%
{\bm\sigma})\right) , \frac{\partial}{\partial {\bf x}}\check{g}_{\nu}\right\} +i [\mathbf{b}\cdot 
{\bm\sigma},\check{g}_{\nu}] \right]  \notag \\
\hspace{-0.4cm}&&=-i[\check{\Sigma}, \check{g}].  \label{eilenberger}
\end{eqnarray}

The self-energy $\check{\Sigma}$ appears in the collision integral on the right hand side and describes
the spin-independent scattering by disorder. The standard self-energy in the
limit of the self-consistent Born approximation has the form~\cite{shen2014}
\begin{equation}
\check{\Sigma}=-\frac{i}{2\tau}\left\langle \check{g} \right\rangle,\qquad%
\frac{1}{\tau}=2\pi n_0 n_i v_{0}^{2},
\label{Born}
\end{equation}
where $n_0=m/2\pi$ is the density of states in the absence of SOC 
(with $m$ being the electron effective mass). The brackets $\langle \cdots \rangle$
denote the angular average over the momentum directions. Finally,  $\tau$
is the elastic scattering time at the Fermi level.

Notice that $\tilde{g}^{R}$ and $\tilde{g}^{A}$ do not depend on the SOC
and, {thus,} have no spin structure, i.e. $\tilde{g}^{R}=\sigma^0$ and $\tilde{g}%
^{A}=-\sigma^0$ solve the retarded and advanced components, respectively,
of Eq.~(\ref{eilenberger}). Then, by using Eq.~(\ref{shift82}), we show
that $g^R=\sigma^0-\partial_{\xi}(\mathbf{b}_0\cdot {\bm\sigma})$. Hence in the 
equilibrium we have 
\begin{eqnarray}
\label{gK}
&&g^K=\tanh\,\left(\frac{\epsilon}{2T}\right)\left(g^R-g^A\right)=
\\  
&&2\tanh\left(\frac{\epsilon}{2T}\right)
\left(\sigma^0-\partial_{\xi}(\mathbf{b}_0\cdot {\bm\sigma})\right)\equiv 
g_{eq}\left[\sigma^0-\partial_{\xi}(\mathbf{b}_0\cdot {\bm\sigma})\right] \notag,
\end{eqnarray}
which defines $g_{eq}$. The Keldysh ($K$) component of the collision integral
can be presented in the form 
\begin{eqnarray}
[\check{\Sigma},\check{g}]^K=\Sigma^{R}g^K+\Sigma^{K}g^A-g^R\Sigma^{K}-g^K\Sigma^{A}.
\end{eqnarray} 
Then the Keldysh components of the linearized Eilenberger equation according
to Eq.~(\ref{eilenberger}) can be written as~\cite{quasiclassical} 
\begin{equation}  \label{keldysh}
(M_0+M_1)g^K=(N_0+N_1) \left\langle g^K \right\rangle,
\end{equation}
where, by defining $\hat{{\bf p}}={\bf p}/|{\bf p}|$, 
\begin{eqnarray}  \label{terms}
&&M_0=g^K+\tau\partial_t g^K +v_F\tau\hat{{\bf p}}\cdot \partial_{\bf x}g^K+i\tau[%
\mathbf{b}_0\cdot{\bm\sigma},g^K],  \label{def_M0} \\
&&\frac{1}{\tau} M_1=-\frac{1}{2}\left\{\frac{\mathbf{b}_0\cdot{\bm\sigma}}{%
p_F}\hat{{\bf p}}-\partial_{\mathbf{p}}(\mathbf{b}_0\cdot{\bm\sigma}%
),\partial_{\bf x} g^K\right\}  \notag \\
&&{\hspace{1.18cm}}-i\left[\partial_{\xi}(\mathbf{b}_0\cdot{\bm\sigma}),\left\{\mathbf{b}_0\cdot%
{\bm\sigma},g^K\right\}\right] \notag\\
&&{\hspace{1.18cm}}- \frac{1}{2\tau}\left\{\partial_{\xi}(\mathbf{b}_0\cdot%
{\bm\sigma}),g^K\right\},  \label{def_M1} \\
&&N_0\left\langle g^K\right\rangle= \left\langle g^K\right\rangle,  \label{def_N0} \\
&&N_1\left\langle g^K\right\rangle=\left\{\partial_{\xi}(\mathbf{b}_0\cdot%
{\bm\sigma}),g^K\right\}.\label{def_N1}
\end{eqnarray}
In the presence of SOC, $\left\langle g^K \right\rangle$ can be written as a
system of four equations according to the spin structure of the
quasiclassical Keldysh Green function, i.e. 
\begin{equation}  \label{gk_spin}
g^K=g_0^K\sigma^0+g_i^K\sigma^i,\qquad i=x,y,z.
\end{equation}

The internal magnetic fields $%
\mathbf{b}=\mathbf{b}_R^{(N)}+\mathbf{b}_D^{(N)}=b_0^{(N)} \hat{\mathbf{b}}^{(N)}$, due to intrinsic RSOC and DSOC can be classified by the power 
$N$ of their momentum dependence~\cite{murkami2006}. 
Notice that we use the notation $(N)$ for the superscript to emphasize  the label character of the symbol $N$ and to avoid confusion with the power function.
In the above equation, $\hat{\mathbf{b}}$ does not depend on
the modulus of the momentum. Hence, the retarded component of the Green
function according to Eqs.~(\ref{g+-})-(\ref{gK}) reads 
\begin{eqnarray}
g^R&=&\sigma^0-c{\hat{\mathbf{b}}^{(N)}}\cdot {{\bm\sigma}}, \qquad c=\frac{Nb_0^{(N)}}{2\epsilon_F},
\end{eqnarray}
where $N=1$ ($N=3$) for the linear  (cubic) SOC. In the presence of both SOCs $\mathbf{b}=\mathbf{b}^{(1)}+\mathbf{b}^{(3)}$. 

We now  consider the Eilenberger equation in the presence of an external
electric field. In order to study an infinite system under a uniform 
time-dependent electric field, we use the minimal substitution 
\begin{equation}  \label{eshift}
\partial_{\mathbf{x}}\to \partial_{\mathbf{x}}-|e| E \hat{\bf E}
\partial_{\epsilon},
\end{equation}
where $|e|$ and $E$ are the absolute values of {the} electron charge and the
applied electric field, respectively, and $\hat{\bf E}\equiv(\hat{E}_x,\hat{E}_y)=(\cos\phi,\sin\phi)$ 
with $\phi$ being the angle of the field with respect to the \textit{x}-axis. Hence,
we can go back to Eq.~(\ref{keldysh}) and solve it for {the} system under
the influence of a uniform time-dependent electric field ${\bf E}=E\hat{\bf E}$ as 
\begin{equation}
M_0g^K=(N_0+N_1)\left\langle g^K \right\rangle +{S}_{\mathbf{E}},
\label{cubicEe}
\end{equation}
from which one obtains
\begin{equation}  \label{nth_gk1}
g^K=M_0^{-1} S_{\mathbf{E}}+M_{0}^{-1}(N_0+N_1) \left\langle
g^K\right\rangle.
\end{equation}
Notice that $g^K$ in Eqs.~(\ref{cubicEe}-\ref{nth_gk1}) represents a column
vector, whose components are defined in Eq.~(\ref{gk_spin}). By taking the
angular average of Eq.~(\ref{nth_gk1}), one obtains a closed equation for $\langle g^K\rangle$ 
in terms of which the physical observables, such as the spin polarizations $S^{i}$
are calculated with~\cite{gorini2010} 
\begin{equation}
S^i=-\frac{n_0}{4}\int_{-\infty}^{\infty} {\rm d}\epsilon \left\langle g_i^K \right\rangle.  \label{spin_relation}
\end{equation}
By using the Pauli matrices expansion of Eq.~(\ref{gk_spin}) in Eq.~(\ref%
{terms}), one can write the Eilenberger equation (\ref{cubicEe}) and (\ref%
{nth_gk1}) as a linear algebraic system for the components $g^K_0$ and $g^K_{i}$. The
explicit matrix form of such a system for Eq.~(\ref{nth_gk1}) is shown in
Appendix \ref{appmatrixform}.
After explicitly taking the average of Eq.~(\ref{nth_gk1})
\begin{equation}  \label{nth_gk2}
(1-\left\langle M_{0}^{-1}(N_0+N_1)\right\rangle)\left\langle g^K\right\rangle=\left\langle M_0^{-1} S_{\mathbf{E}}\right\rangle,
\end{equation}
we can neglect $N_1$ to leading order in $b_0/\epsilon_F$, thus decoupling the spin sector from the charge one. Furthermore, by using the expressions of $M_0$, $N_0$ and $S_{\bf E}$ from the Eqs.(\ref{m0}, \ref{n0n1}, \ref{termsn1}) one can show that after angular average $g_z$ decouples from $g_x$ and $g_y$. As a result the in-plane spin dynamics 
is reduced to problems described
by $2\times 2$ matrices.

\section{Inverse spin-galvanic effect: beyond the diffusive regime}

In this section we will evaluate the ISGE in the presence of the linear RSOC
and DSOC. The evaluation will not be restricted  to the diffusive approximation 
$b\tau \ll 1$, where the SOC is small compared to the disorder broadening.
Hence, we will extend the previous results obtained in the diffusive 
regime~\cite{roberto_ann2012,giglberger2007,vignale2014,shelankov2010,malekiinverse,paper1,paper2_physB,
shen2014microscopic,burkov2004theory,schwab2010inverse,smirnov2017electrical}.
This case will also serve as an example of the way our formalism works. In a
2DEG the effective magnetic field due to the combination of the linear RSOC
and DSOC reads~\cite{Winkler2003} 
\begin{equation}
\mathbf{b}^{(1)}=p%
\begin{bmatrix}
\alpha_1 \hat{p}_{y}+\beta_1 \hat{p}_{x} \\ 
-\alpha_1 \hat{p}_{x}-\beta_1 \hat{p}_{y} \\ 
0%
\end{bmatrix}
,
\end{equation}%
where $\alpha_1 $ and $\beta_1 $ are the magnitudes of the linear RSOC and
DSOC, respectively. 
The terms $S_{\mathbf{E}}$ proportional to the uniform electric
field are derived by using Eq.~(\ref{termsn1}) 
\begin{equation}
S_{\mathbf{E}}=\tilde{E}%
\begin{bmatrix}
s_{11} & s_{12} \\ 
s_{21} & s_{22}%
\end{bmatrix}%
\begin{bmatrix}
\hat{E}_{x} \\ 
\hat{E}_{y}%
\end{bmatrix}\label{Se_form}%
\end{equation}%
with $\tilde{E}=-|e|E\tau  v_F \partial_{\epsilon }g_{eq}$ and 
\begin{align}
{}& s_{11}=\alpha_1 \sin 2\phi +\beta_1 \cos 2\phi,  \notag \\
{}& s_{12}=-\alpha_1 \cos 2\phi +\beta_1 \sin 2\phi,  \notag \\
{}& s_{21}=-\alpha_1 \cos 2\phi -\beta_1 \sin 2\phi,  \notag \\
{}& s_{22}=-\alpha_1 \sin 2\phi +\beta_1 \cos 2\phi.  \label{sij}
\end{align}%

By performing the angular average of Eq.~(\ref{nth_gk1}) for a uniform system, one gets 
\begin{equation}
\hat{\Gamma}\left\langle g^{K}\right\rangle =\left\langle M_{0}^{-1}S_{%
\mathbf{E}}\right\rangle,
\label{ekeldysh}
\end{equation}%
where $\hat{\Gamma}=1-\left\langle M_{0}^{-1}(N_{0}+N_{1})\right\rangle $
includes both the spin relaxation and the frequency dependence effects. To solve the above
equation, we have to perform several integrals with respect to the
momentum direction, as listed in Appendix~\ref{integrall}. Under the
uniform time-dependent electric field, we have 
\begin{equation}
\left\langle M_{0}^{-1}(N_{0}+N_{1})\right\rangle =\frac{1}{%
L^{3}+La^{2}(\alpha_1 ^{2}+\beta_1 ^{2})}\left[ 
\begin{array}{cc}
M_{11} & M_{12} \\ 
M_{21} & M_{22}%
\end{array}%
\right] , \label{matrixM}
\end{equation}%
where $a=2\tau p_{F}$ and $L=1-i\tau \Omega $, with $\Omega$ the 
variable associated to the Fourier transform with respect to time $t$ (see the term with the time derivative in Eq.(\ref{def_M0})) . The matrix elements appearing in Eq.(\ref{matrixM}) read
\begin{eqnarray}
M_{11}=M_{22} &=&\left[L^{2}+\frac{a^{2}}{2}(\alpha_1 ^{2}+\beta_1 ^{2})\right] 
\frac{1}{\sqrt{1-\mathcal{C}^{2}}}  \notag \\
&&-\frac{a^{2}\alpha_1 \beta_1 }{\mathcal{C}}\frac{1-\sqrt{1-\mathcal{C}^{2}}}{%
\sqrt{1-\mathcal{C}^{2}}},  \label{m11} \\
M_{12}=M_{21} &=&a^{2}\frac{\alpha_1 ^{2}+\beta_1 ^{2}}{2\mathcal{C}}\frac{1-%
\sqrt{1-\mathcal{C}^{2}}}{\sqrt{1-\mathcal{C}^{2}}}  \notag \\
&&-a^{2}\frac{\alpha_1 \beta_1 }{\sqrt{1-\mathcal{C}^{2}}},  \label{m12}
\end{eqnarray}%
with 
\begin{equation}
\mathcal{C}=a^{2}\frac{2\alpha_1 \beta_1 L}{L^{3}+La^{2}(\alpha_1 ^{2}+\beta_1^{2})}%
.  \label{Cterm}
\end{equation}%
For the two dimensionless quantities $a\alpha_1 $ and $a\beta_1 $ we may consider two different regimes. As we
assumed at the beginning, the SO splitting and the disorder broadening are
much smaller than the Fermi energy $\epsilon _{F}$. For instance, in the
Rashba model we have 
\begin{equation}
\epsilon _{F}\gg \frac{1}{\tau },\qquad \epsilon _{F}\gg 2\alpha_1\, p_{F}.
\label{splitting_broadening}
\end{equation}%
We can rewrite $a\alpha_1 $ in terms of the two small parameters $\alpha_1
/v_{F} $ and $1/\epsilon _{F}\tau $ as 
\begin{equation}
a\alpha_1 =2\tau\alpha_1\, p_{F}=\frac{4\alpha_1 }{v_{F}}\epsilon _{F}\,\tau .
\end{equation}%
According to Eq.~(\ref{splitting_broadening}) and the relation between $%
\alpha_1 /v_{F}$ and $1/\epsilon _{F}\tau $ one can define two different regimes
depending on which one dominates~\cite{halperin2004,khaetskii2006}. 
The first one is the diffusive regime, corresponding to a high impurity concentration,
i.e. $a\alpha_1 \ll 1$ and the Dyakonov-Perel spin relaxation. The second regime 
which occurs at $a\alpha_1 \gg 1$, goes beyond the diffusive limit and describes the opposite situation of 
a relatively low concentration of impurities, 
where the spin relaxation time is close to $\tau$~\cite{gridnev2001}. 
To analyze these two regimes and a crossover between them in a simple form, we focus 
in this Section on a model with the linear RSOC and DSOC only. 
In the diffusive limit ($\mathcal{C}\rightarrow
0 $) we can neglect the terms with higher order of the Rashba-Dresselhaus SOC
since 
\begin{equation}
\frac{\sqrt{1+\mathcal{C}^{2}}-1}{\sqrt{1+\mathcal{C}^{2}}}\approx\frac{\mathcal{C}^{2}}{2}\ll 1.
\end{equation}%
In such a case, the second term in Eq.~(\ref{m11}) and the first one in Eq.~(%
\ref{m12}) vanish. One may notice also that $\mathcal{C}=0$ when 
either $\alpha_1=0$ or $\beta_1=0$.
Finally, by using Eq.~(\ref{spin_relation}) we can write a generalized Bloch equation for the spin density as
\begin{equation}
\hat{\Gamma}\mathbf{S}=\hat{\omega} \hat {\mathbf{E}},
\label{linear_spin}
\end{equation}%
where the matrix $\hat{\omega}$ describing the spin generation torque in
the right hand side of the above equation is given by 
\begin{equation}
\hat{\omega}=\frac{S_{0}}{2}\frac{\beta_1 ^{2}-\alpha_1 ^{2}}
{L^{3}+La^{2}(\alpha_1 ^{2}+\beta_1 ^{2})}
\begin{bmatrix}
\omega _{11} & \omega _{12} \\ 
\omega _{21} & \omega _{22}%
\end{bmatrix}
,
\end{equation}%
with $S_{0}=-|e|\tau n_{0}E$ and 
\begin{eqnarray}
\omega _{11}=-\omega _{22} &=&-\frac{\beta_1 a^{2}}{\sqrt{1-\mathcal{C}^{2}}}%
-\alpha_1 \delta, \\
\omega _{12}=-\omega _{21} &=&\frac{\alpha_1 a^{2}}{\sqrt{1-\mathcal{C}^{2}}}%
+\beta_1 \delta,
\end{eqnarray}%
and 
\begin{equation}
\delta =L^{2}\frac{1}{(\alpha_1 ^{2}+\beta_1 ^{2})\mathcal{C}-2\alpha_1 \beta_1 }%
\frac{1-\sqrt{1-\mathcal{C}^{2}}}{\sqrt{1-\mathcal{C}^{2}}}.
\end{equation}%
Correspondingly, $\hat{\Gamma}$ has the phenomenological meaning of a spin relaxation torque matrix,
and the resulting spin density $\mathbf{S}$ is obtained as a result of the balance between
the generation and the relaxation torques.  
Clearly, in the diffusive regime, when $\mathcal{C}\ll 1$, $\delta$ is very small,
and one recovers the standard DP spin relaxation. Equation (\ref{linear_spin}) is
one of the main results of the paper. The contributions to the spin torque, dependent 
on $\delta $ and $\mathcal{C}$, appear in Eq. (\ref{linear_spin}) only
when the interplay of the Rashba-Dresselhaus SOC is considered beyond the conventional
diffusive approximation of Refs.~\cite{paper2_physB,malekiinverse,paper1}. 
Furthermore, the powers of $L$ take into account terms relevant at high
frequencies. We also notice that at $\alpha_{1}^{2}=\beta_{1}^{2}$, the contributions of the RSOC and the
DSOC cancel each other, which leads to a pure gauge configuration ~\cite{tokatly2010}, 
where the CISP does not appear. 

\subsection{Inverse spin galvanic effect in the linear Rashba model}

\label{rashba_model} In this subsection, we will solve the generalized Bloch equations~(%
\ref{linear_spin}) numerically for different RSOC magnitudes. 
After setting $\beta_{1}=0$ in Eq.~(\ref{linear_spin}), the Bloch
equations in the 2DEG Rashba model read 
\begin{eqnarray}
\begin{bmatrix}
S^x \\ 
S^y%
\end{bmatrix}%
= \frac{1}{2}\frac{S_0^{\alpha}\alpha_1^2 a^2 E}{L^3-L^2+\left(L-1/2\right)a^2\alpha_1^2} 
\begin{bmatrix}
\hat{E}_y \\ 
\hat{E}_x%
\end{bmatrix}%
,
  \label{eerashba}
\end{eqnarray}
with $S_0^{\alpha}=-|e|n_0\tau \alpha_1$. 
In the static limit when the frequency is zero,
i.e. $L=1$, the spin polarization becomes 
\begin{eqnarray}  \label{EE}
\begin{bmatrix}
S^x \\ 
S^y%
\end{bmatrix}%
=S_0^{\alpha} E 
\begin{bmatrix}
\hat{E}_y \\ 
\hat{E}_x%
\end{bmatrix}%
,  \label{rashbal0}
\end{eqnarray}
which is the Edelstein result~\cite{Edelstein1990}. In the following equation, we consider the
frequency-dependent ISGE by inserting $L=1-i\Omega \tau$ in Eq.~(\ref{eerashba}).  
In this case, the real and imaginary components of the spin density become zero, respectively,
when 
\begin{eqnarray}
\Omega \tau&=& \frac{a\alpha_1}{2},\nonfrenchspacing \label{s0r}\\
\Omega \tau&=& 0;\,\,\, \sqrt{1+a^2\alpha_1^2} .  \label{s0}
\end{eqnarray}

\begin{figure}
	\centering
	\begin{tabular}{@{}p{\linewidth}@{\qquad}p{\linewidth}@{}@{\qquad}p{\linewidth}@{}}
		\subfigimg[width=\linewidth,pos=ur]{(a)}{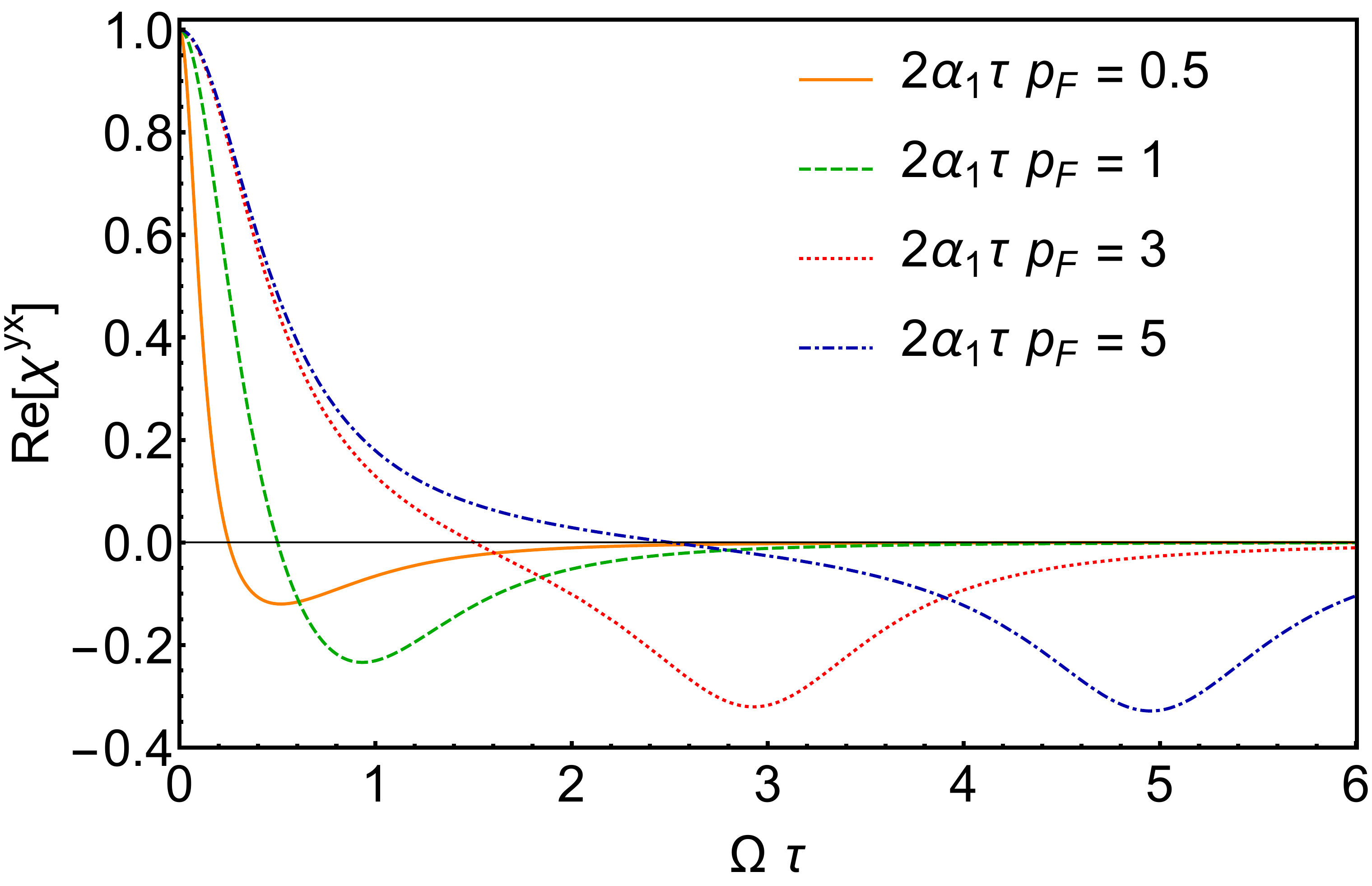} \\
		\subfigimg[width=\linewidth,pos=ur]{(b)}{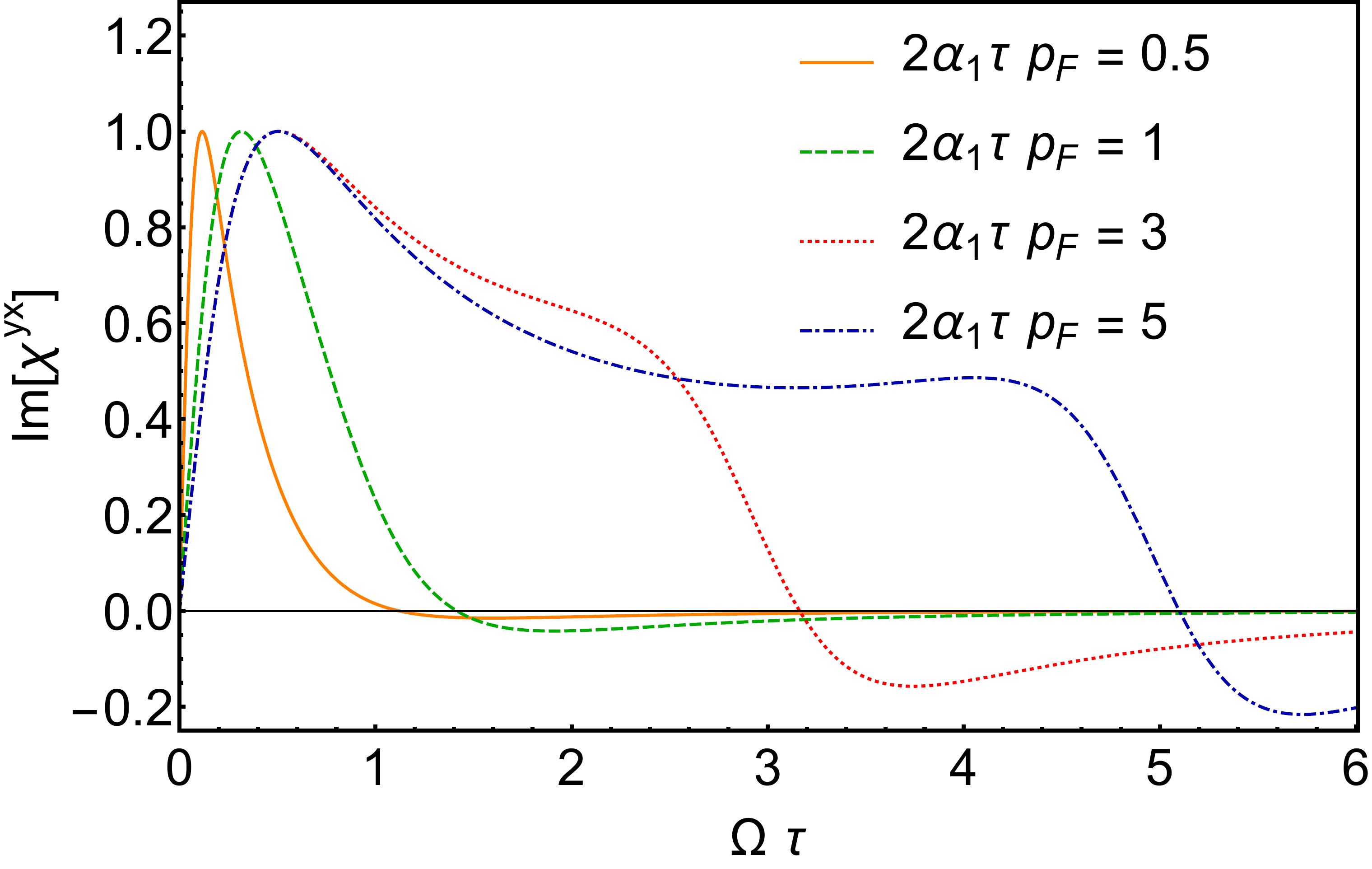} \\
		\subfigimg[width=\linewidth,pos=ur]{(c)}{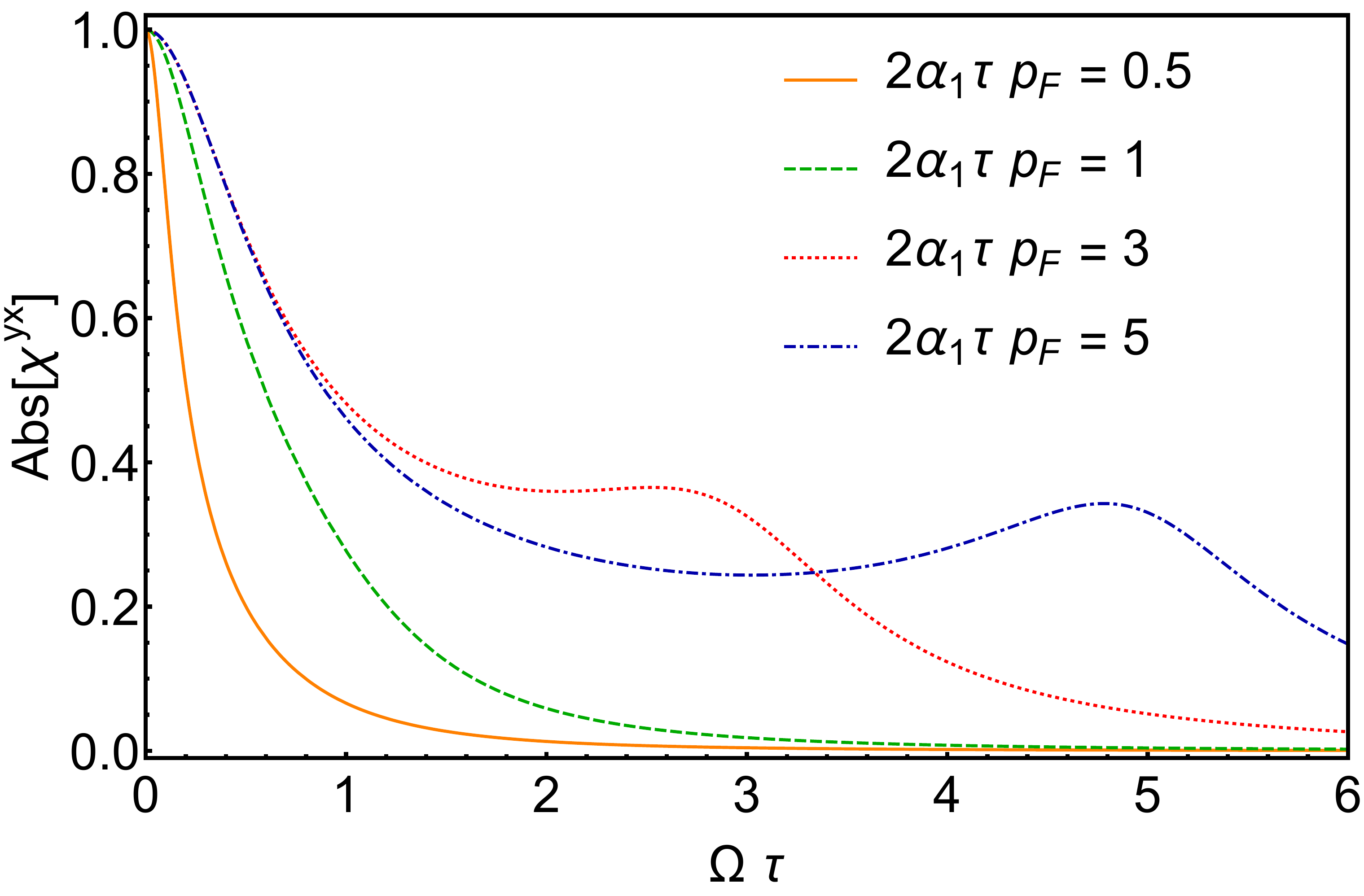}
	\end{tabular}
	\caption{(a) Real part, (b) imaginary part and (c) absolute value of the normalized SG conductivity $\protect\chi^{yx}$ as a function of the frequency $\Omega\tau$.  In all plots: $2\alpha_1\tau p_F=0.5$ (solid orange), and $2\alpha_1\tau p_F=1$ (dashed green), and $2\alpha_1\tau p_F=3$ (dotted red),
		and $2\alpha_1\tau p_F=5$ (dot-dashed blue). Results are given in units of $S_{0}^{\alpha}$.}
	\label{riEE_Fa}
\end{figure}

When the imaginary part of the ISGE vanishes, the real part dominates, and
vice versa, leading to a dependence of the ISGE on the field frequency. We define the frequency-dependent
spin-galvanic (SG) conductivity, which can be found from Eq.~(\ref{eerashba}), as 
\begin{equation}  \label{eef0}
S^i(\Omega)=\chi^{ij}_{SG}(\Omega)\, E_j(\Omega),\qquad i,j=x,y.
\end{equation}
Notice that both the charge current and spin density are odd under time reversal. Hence the Onsager relations imply that  the SG and inverse SG conductivities are equal.
In the numerics
we use the normalized conductivities 
\begin{equation}
\chi^{ij}=\frac{\chi_{SG}^{ij}(\Omega)}{\chi_{SG}^{ij}(\Omega_{\max})},
\end{equation}
to evaluate their frequency behavior, $\Omega_{\max}$ being the frequency at the maximum $\chi_{SG}^{ij}$. In Fig.~%
\ref{riEE_Fa}, we plot the real and imaginary parts as well as the absolute value of
the normalized conductivity $\chi^{yx}$ as a function of the frequency
in  units of $S_0^{\alpha}$ for the different magnitudes of
RSOC. At sufficiently high frequencies, the conductivity vanishes, according to Eqs.~(\ref{s0r})-(\ref{s0}),
and a significant conductivity oscillation appears 
at $\Omega \sim \alpha_{1}p_{F}$ if one goes beyond the diffusive regime, that is the 
condition $\alpha_{1}p_{F}\tau\agt 1$ is satisfied. 

\begin{figure*}
	\centering
	\begin{tabular}{@{}p{0.48\linewidth}@{\qquad}p{0.48\linewidth}@{}@{}p{0.48\linewidth}@{\qquad}p{0.48\linewidth}@{}}
		\subfigimg[width=\linewidth,pos=ur]{(a)}{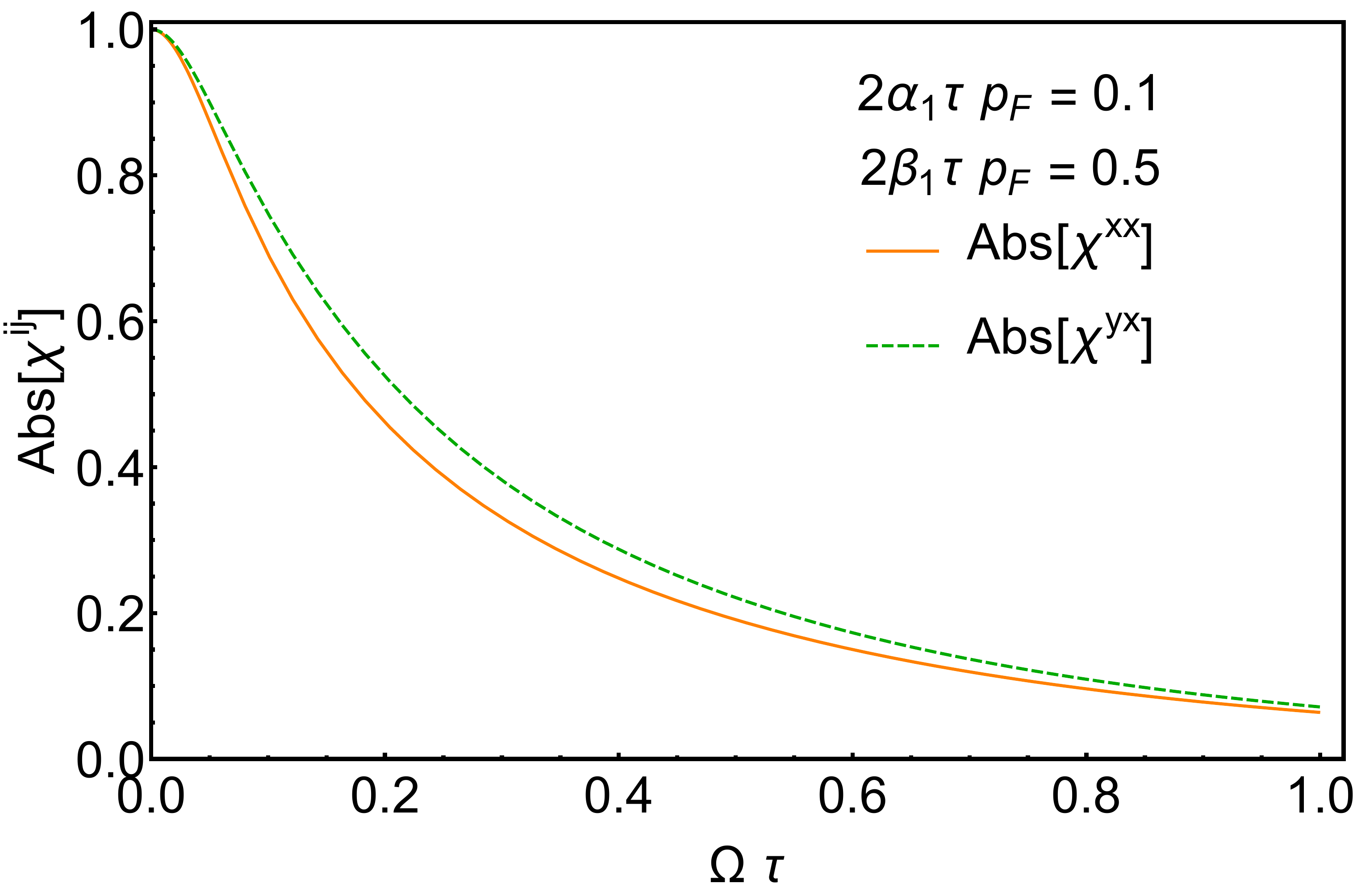} &
		\subfigimg[width=\linewidth,pos=ur]{(b)}{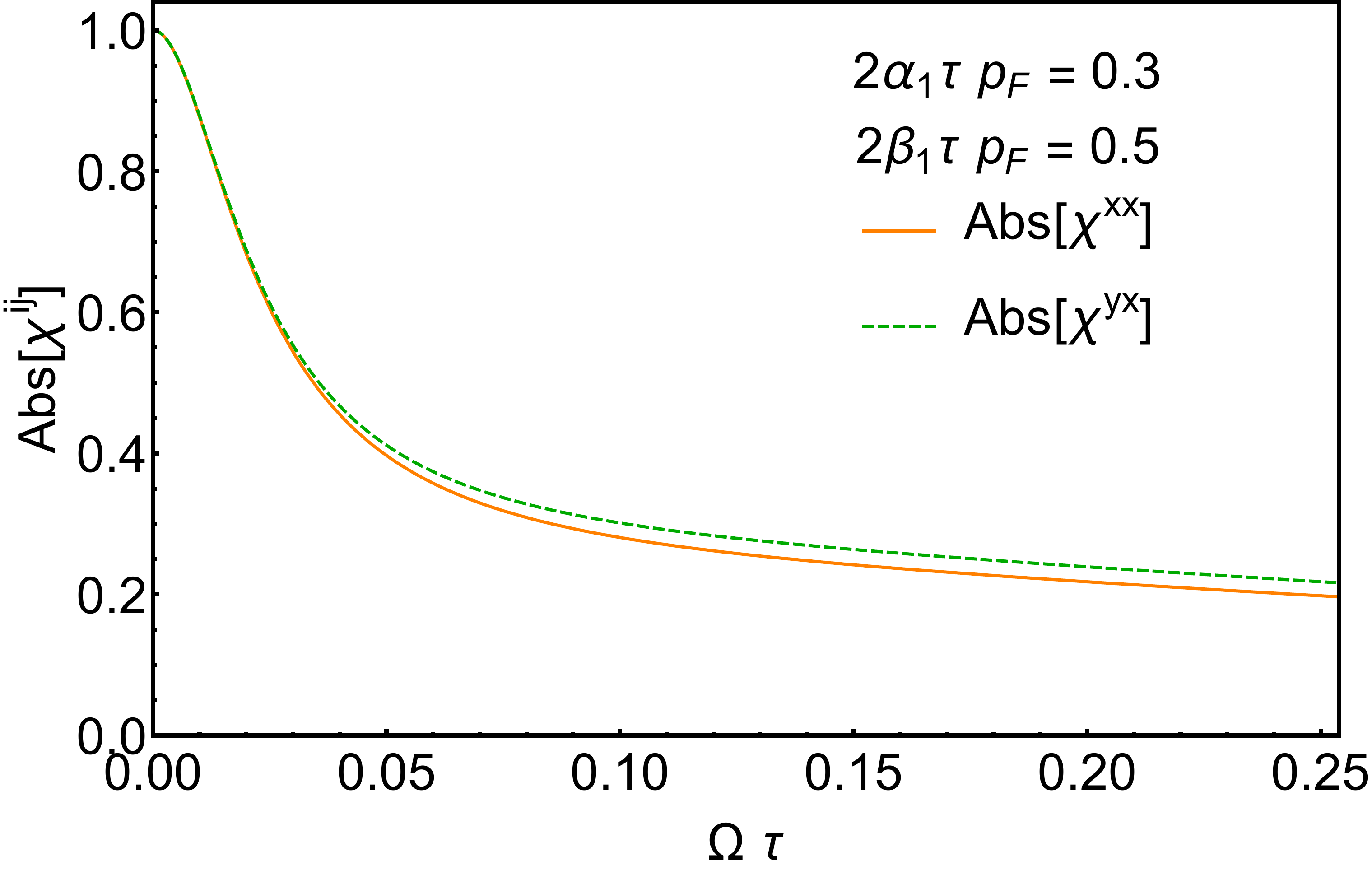} \\
		\subfigimg[width=\linewidth,pos=ur]{(c)}{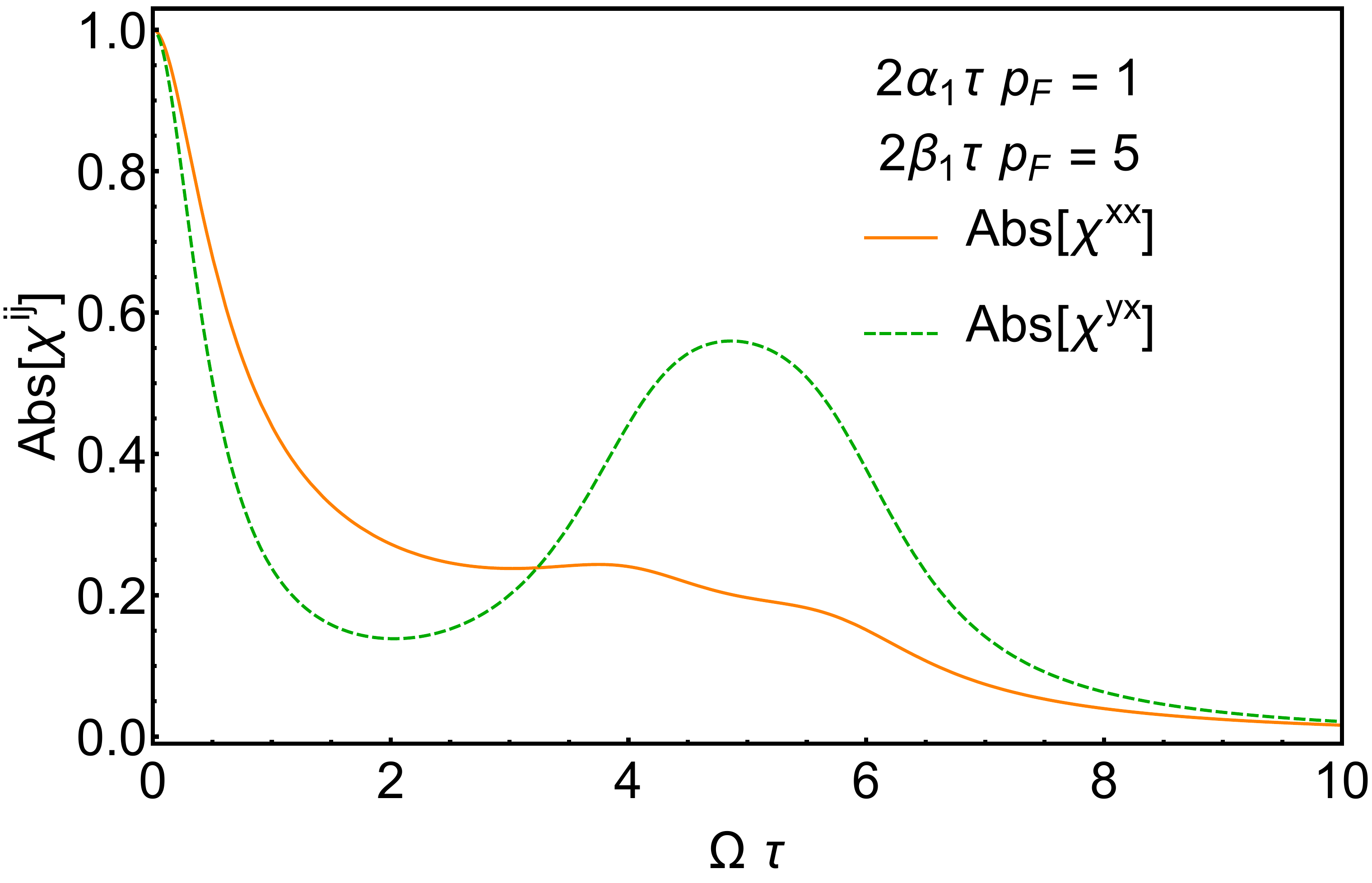} &
		\subfigimg[width=\linewidth,pos=ur]{(d)}{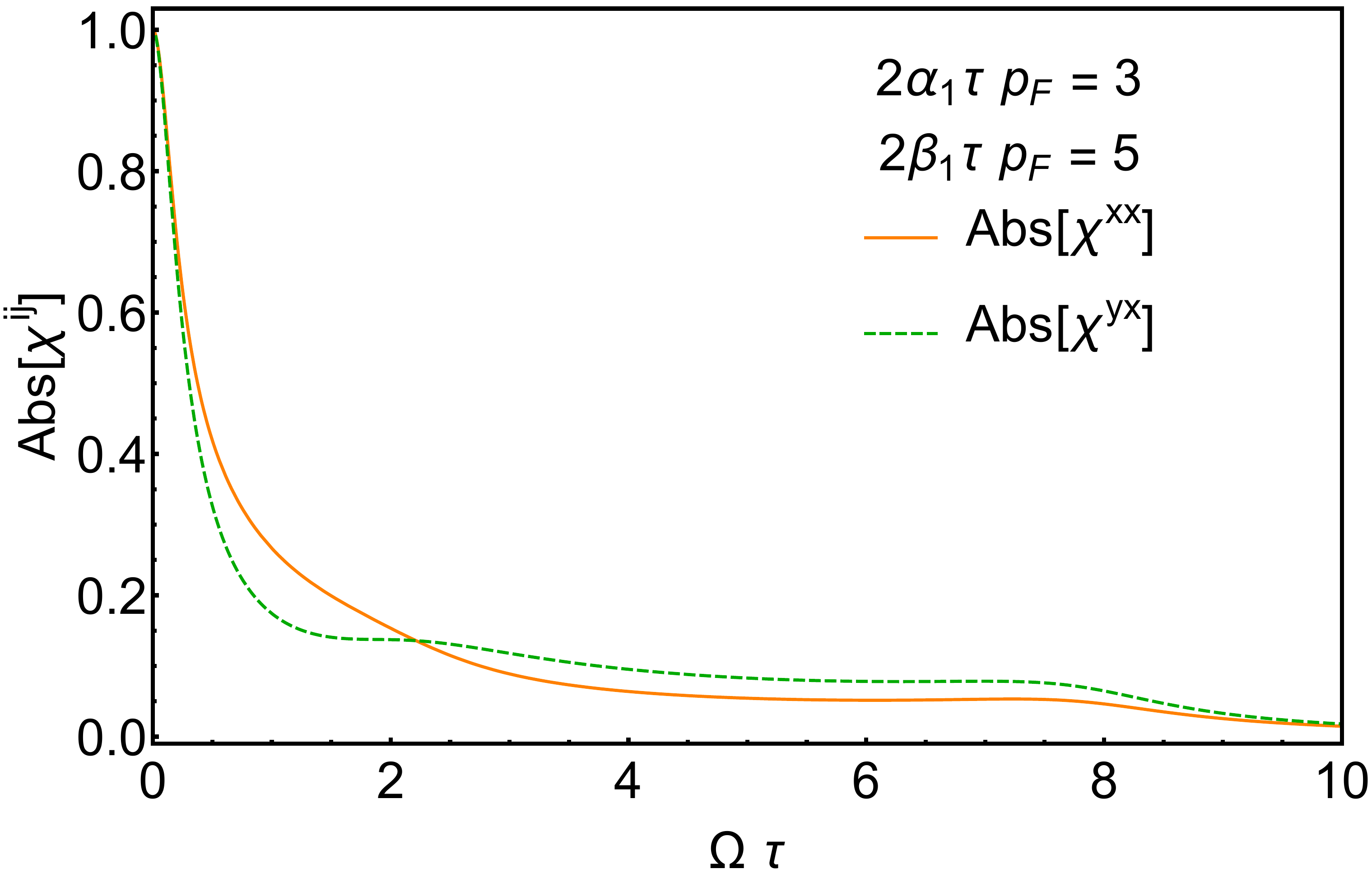}
	\end{tabular}
		\caption{Plots of the absolute value of the normalized SG conductivity ($\chi^{ij}=\chi^{xx}$ 
			(solid orange): $\chi^{yx}$(dashed green)) as a function of frequency in the presence of linear RSOC and DSOC. 
			From the left to the right: (a), (b) conductivity in the diffusive regime and (c), (d) conductivity beyond the 
			diffusive regime. The linear SOC coefficients from the top to the bottom: 
			(a) $2\alpha_1\tau p_F=0.1$ and $2\beta_1\tau p_F=0.5$; (c) $2\alpha_1\tau p_F=1$ 
			and $2\beta_1\tau p_F=5$; (b) $2\alpha_1\tau p_F=0.3$ and $2\beta_1\tau p_F=0.5$; 
			(d) $2\alpha_1\tau p_F=3$ and $2\beta_1\tau p_F=5$. The results are given in units of $S_0^{\protect\alpha}$.}
		\label{diffusivesxy}
\end{figure*}

\subsection{Inverse spin-galvanic effect in the linear Rashba-Dresselhaus
SOC}

As shown in the previous subsection, the ISGE shows a  different behavior
with respect to the dimensionless parameter $a\alpha_1$. In this subsection we consider
the ISGE in the presence of both the RSOC and DSOC. In the diffusive regime
we assume $a \alpha_1\ll 1$ and $a \beta_1\ll 1$ for high impurity concentration. 
In this limit, we can neglect the higher order terms in the 
Rashba-Dresselhaus SOC and in $\Omega \tau$. Hence, the generalized Bloch equation 
has the same form as Eq.~(\ref{linear_spin}) with the $\hat{\Gamma}$ and $\hat{\omega%
}$ given by
\begin{eqnarray}
\hat{\Gamma}&=&-i\Omega\tau+\frac{a^2}{2} 
\begin{bmatrix}
\alpha_1^2+\beta_1^2 & 2\alpha_1 \beta_1 \\ 
2\alpha_1 \beta_1 & \alpha_1^2+\beta_1^2 
\end{bmatrix}%
\label{49}
, \\
\hat{\omega}&=&S_0\frac{a^2}{2}(\beta_1^2-\alpha_1^2)%
\begin{bmatrix}
-\beta_1 & \alpha_1 \\ 
-\alpha_1 & \beta_1%
\end{bmatrix}%
.
\label{50}
\end{eqnarray}
Notice that in the diffusive approximation,  keeping just the first order in $%
\Omega \tau$ yields the standard form of the Bloch equation, which coincides with the result of Refs.~\cite{malekiinverse,paper2_physB,paper1}, when the extrinsic effect
is not considered. 
For $\beta_1=0$, the above equations reproduce the results for the Rashba
model presented in Eq.~(\ref{eerashba}) for the diffusive limit.
In the limit of the spin helix regime, where RSOC and DSOC are close to each other, we can write the spin polarization of Eq.~(\ref{linear_spin}) as
	\begin{eqnarray}
	\begin{bmatrix}
	S^x(\Omega)\\
	S^y(\Omega)
	\end{bmatrix}=S_0^{\alpha} \frac{a^2\Delta^2E}{-2i\Omega\tau+a^2\Delta^2}
	\begin{bmatrix}
	 \hat{E}_x+\hat{E}_y\\
	 -\hat{E}_x- \hat{E}_y
	\end{bmatrix},
	\label{helicity}
	\end{eqnarray}
where $\Delta=\alpha_1-\beta_1$ and $|\Delta|\ll|\alpha_{1}|$, $\Omega \tau\ll1$.
We notice that there is no effect for $\Delta =0$, as expected, and the typical frequency scale is $\Omega\sim a^2\Delta^2/2\tau$. 
This can be appreciated explicitly by the numerical evaluation of the
SG conductivities  when the RSOC and DSOC are present.
 In Fig.~\ref{diffusivesxy}(a-b), we plot the normalized conductivities, $\chi^{xx}$ and $\chi^{xy}$, as function of the frequency  
for different values of $\alpha_1$ and $\beta_1$ in the diffusive regime.
The different scale in the frequency behavior from the top to the bottom plots is related to the difference 
between the two RSOC and DSOC as shown in Eq.~(\ref{helicity}).
In the diffusive regime, there is no finite-frequency peak in the conductivity, independent of the spin-orbit coupling details.

To solve numerically the generalized Bloch equations beyond the diffusive approximation, we have
to keep all the orders of the spin-orbit coupling field $\mathbf{b}$ and frequencies $\Omega \tau$. As we demonstrated 
in Eq.~(\ref{linear_spin}), in this regime several new terms in the
spin relaxation and the spin generation torques contribute to the
Bloch equation. Figures ~\ref{diffusivesxy}(c-d) show the
numerically obtained absolute value of the SG conductivity  as a function of frequency 
beyond the diffusive regime. In contrast to the diffusive regime, we
find a finite-$\Omega$ SG conductivity. The latter increases with the difference of the magnitudes of RSOC and DSOC.

\section{Inverse Spin-Galvanic Effect in the cubic Rashba-Dresselhaus Model}

For a quantum well with the cubic Rashba-Dresselhaus SOC, the
Hamiltonian contains $p$-cubic contribution in addition to the $p$-linear terms~%
\cite{pikus84,cubicSOC}. According to Eq.~(2) of Ref.~\cite{cubicrashba},
the effective Hamiltonian of the structural inverse asymmetry to the third
order in the wave vector $\mathbf{p}$ reads 
\begin{equation}
H_{R}^{(3)}=i\alpha _{3}%
\begin{bmatrix}
0 & {(p_{x}-ip_{y})}^{3} \\ 
-(p_{x}+ip_{y})^{3} & 0%
\end{bmatrix}%
\equiv\mathbf{b}_{R}^{(3)}\cdot{\bm\sigma}
\label{cubicR}
\end{equation}%
with ${\bf b}_{R}^{(3)}$ being the effective internal magnetic field due to the cubic Rashba
SOC, which can also be written as:
\begin{equation}
\mathbf{b}_{R}^{(3)}=
\alpha _{3}
\begin{bmatrix}
3{\Large p_{y}\,p_{x}^{2}-p_{y}^{3}} \\ 
3p_{x}\,p_{y}^{2}-p_{x}^{3}
\end{bmatrix}%
=\alpha _{3}p^3
\begin{bmatrix}
\sin 3\phi \\ 
-\cos 3\phi 
\end{bmatrix}%
.
\label{cubic_rashba}
\end{equation}%
In quantum wells,
the Hamiltonian also contains the terms arising due to the
bulk inversion asymmetry, i.e. the cubic Dresselhaus SOC~\cite{cubicSOC} 
\begin{equation}
H_{D}^{(3)}=
-\beta _{3}
\begin{bmatrix}
0 & (p_{x}-ip_{y})^{3} \\ 
(p_{x}+ip_{y})^{3}& 0%
\end{bmatrix}%
\equiv\mathbf{b}_{D}^{(3)}\cdot {\bm\sigma}
\end{equation}%
or, alternatively, 
\begin{equation}
\mathbf{b}_{D}^{(3)}=
\beta _{3}
\begin{bmatrix}
3p_{x}\,p_{y}^{2}-p_{x}^{3} \\ 
-(3p_{y}\,p_{x}^{2}-p_{y}^{3})
\end{bmatrix}=%
-\beta _{3}p^3
\begin{bmatrix}
\cos 3\phi \\ 
\sin 3\phi 
\end{bmatrix}%
.
\end{equation}%
Hence, the total effective internal magnetic field of the cubic
Rashba-Dresselhaus SOC is given by~\cite{Winkler2003} 
\begin{eqnarray}
\mathbf{b}^{(3)} &=&\mathbf{b}_{R}^{(3)}+\mathbf{b}_{D}^{(3)}=p^3
\begin{bmatrix}
\alpha _{3}\sin  3\phi -\beta _{3}\cos 3\phi \\ 
-\alpha _{3}\cos  3\phi -\beta _{3}\sin 3\phi%
\end{bmatrix}
\notag \\
&\equiv&b_{0}^{(3)}\hat{\mathbf{b}}^{(3)}.
\end{eqnarray}%
To the linear order in the external electric field, the source term $S_{\mathbf{E}}$
has the same form as in Eq.~(\ref{Se_form}) with Eq.~(\ref{sij}) replaced by 
\begin{align}
& {}s_{11}=p_F^{2}\alpha _{3}\left( 2\sin 4\phi -\sin 2\phi \right)
+p_F^{2}\beta _{3}\left( -2\cos 4\phi +\cos 2\phi \right),  \notag \\
& {}s_{21}=p_F^{2}\alpha _{3}\left( 2\cos 4\phi +\cos 2\phi \right)
+p_F^{2}\beta _{3}\left( -2\sin 4\phi +\sin 2\phi \right),  \notag \\
& {}s_{12}=-p_F^{2}\alpha _{3}\left( 2\cos 4\phi +\cos 2\phi \right)
-p_F^{2}\beta _{3}\left( 2\sin 4\phi +\sin 2\phi \right),  \notag \\
& {}s_{22}=-p_F^{2}\alpha _{3}\left( 2\sin 4\phi +\sin 2\phi \right)
+p_F^{2}\beta _{3}\left( 2\cos 4\phi +\cos 2\phi \right).
\end{align}%
By using Eqs.~(\ref{m0})-(\ref{termsn1}) one obtains for the generalized Bloch equation:
\begin{eqnarray}
\hat{\Gamma} &=&\frac{L^{2}+\frac{1}{2}a^{2}p_F^{4}(\alpha_{3}^{2}+\beta_{3}^{2})}{L^{3}+La^{2}p_F^{4}
(\alpha_{3}^{2}+\beta _{3}^{2})}\sigma_{0}, \\
\hat{\omega} &=&0.
\end{eqnarray}%
The above equations show that, in the cubic RSOC and DSOC model, the spin
generation torque $\hat{\omega}\mathbf{E}$ vanishes, although the spin
relaxation rate $\hat{\Gamma}$ is nonzero. Notice that this result can also
be derived in the diagrammatic approach, where it appears as a consequence
of the vanishing of the vertex corrections. The latter contain the first harmonics of $\phi$ and, 
hence, the ${\bf b}$ field with the third harmonics does not contribute. 
This was first noticed by Murakami in the theory of the spin Hall effect~\cite{murakami_2004}. 

\section{Inverse spin galvanic effect in the linear and cubic
Rashba model}

As we have seen in the previous section, in the 
cubic SOC case, the ISGE does not exist. Here we evaluate the  ISGE in the presence of both the  linear and cubic RSOCs.
In this case, the internal magnetic field reads 
\begin{equation}
\mathbf{b}_{\mathcal{R}}=p%
\begin{bmatrix}
\alpha _{1}\sin\phi +\alpha _{3}\,p^2\sin  3\phi \\ 
-\alpha _{1}\cos\phi -\alpha _{3}\,p^2\cos  3\phi\\ 
\end{bmatrix}%
,
\end{equation}%
where $\alpha _{1}$ and $\alpha _{3}$ are the magnitudes of the linear and
cubic Rashba SOC, respectively. With the field $\mathbf{b}_{\mathcal{R}}$ in Eq.~(\ref{termsn1}), 
the source $S_{\mathbf{E}}$ in Eq.~(\ref{Se_form}) becomes: 
\begin{align}
& {}s_{11}=2p_F^{2}\,\alpha _{3}\sin 4\phi +(\alpha _{1}-p_F^{2}\alpha _{3})\sin
2\phi,  \notag \\
& {}s_{21}=-2p_F^{2}\,\alpha _{3}\cos 4\phi +(-\alpha _{1}+p_F^{2}\alpha _{3})\cos
2\phi,  \notag \\
& {}s_{12}=-2p_F^{2}\,\alpha _{3}\cos 4\phi -(\alpha _{1}+p_F^{2}\alpha _{3})\cos
2\phi,  \notag \\
& {}s_{22}=-2p_F^{2}\,\alpha _{3}\sin 4\phi -(\alpha _{1}+p_F^{2}\alpha _{3})\sin
2\phi .  \label{termsn}
\end{align}%
By using Eqs.~(\ref{m0}) and (\ref{n0n1}), the matrix $\left\langle
M_{0}(N_{0}+N_{1})\right\rangle $ can be written as 
\begin{equation}
\left\langle M_{0}(N_{0}+N_{1})\right\rangle =\frac{1}{%
L^{3}+L(a_{1}^{2}+a_{3}^{2})}%
\begin{bmatrix}
M_{11} & 0 \\ 
0 & M_{22}  
\end{bmatrix}%
,
\end{equation}%
with $a_{1}=a\alpha _{1}$ and $a_{3}=ap_{F}^{2}\,\alpha _{3}$ and 
\begin{eqnarray}
M_{11} &=&(L^{2}+\frac{1}{2}(a_{1}^{2}+a_{3}^{2}))A_{0}  \label{mrlc1} \\
&+&\frac{1}{2}(-a_{1}^{2}+2a_{1}a_{3})A_{2}-a_{1}a_{3}A_{4}-\frac{1}{2}%
a_{3}^{2}A_{6},  \notag \\
M_{22} &=&(L^{2}+\frac{1}{2}(a_{1}^{2}+a_{3}^{2}))A_{0}  \label{mrlc2} \\
&+&\frac{1}{2}(a_{1}^{2}+2a_{1}a_{3})A_{2}+a_{1}a_{3}A_{4}+\frac{1}{2}%
a_{3}^{2}A_{6},  \notag
\end{eqnarray}%
where%
\begin{equation}
A_{n}=\left\langle \frac{\cos (n\phi )}{1+\mathcal{D}\cos 2\phi }%
\right\rangle,
\end{equation}%
and 
\begin{equation}
\mathcal{D}=\frac{2La_{1}a_{3}}{L^{3}+L(a_{1}^{2}+a_{3}^{2})}.
\end{equation}%
The diffusive regime occurs when $a_{1}\ll 1$ and $a_{3}\ll 1,$ and for this
regime one has $\mathcal{D}\ll 1$. In such case, all the integrals except the
first one in Eqs.~(\ref{mrlc1}) and (\ref{mrlc2}) can be neglected. Moreover $\mathcal{D}=0$
when either $a_{1}=0$ or $a_{3}=0$.

Finally, by using Eqs.~(\ref{m0}) and (\ref{termsn1}) the matrix $\hat{\omega}$ appearing on the
right hand side of Eq.~(\ref{linear_spin}) can be presented as: 
\begin{equation}
\hat{\omega}=\frac{S_{0}}{L^{3}+L(a_{1}^{2}+a_{3}^{2})}%
\begin{bmatrix}
0 & w_{12} \\ 
w_{21} & 0 
\label{w12w21}
\end{bmatrix}%
,
\end{equation}%
with  
\begin{eqnarray}
&&\hspace{-0.1cm} w_{12}=\frac{\alpha _{1}}{2}\Big(a_{1}^{2}+3a_{3}^{2}\Big)A_{0}
\label{wlc} \nonumber \\
&&\hspace{-0.6cm}+\left[-L^{2}(\alpha _{1}+p_{F}^{2}\alpha _{3})+\frac{\alpha _{1}}{2}%
(a_{1}^{2}+2a_{3}^{2}+6a_{1}a_{3}) +2\alpha _{3}p_F^{2}a_{3}^{2}\right]A_{2}  \notag \\
&+&\frac{1}{4}\left(\alpha _{1}a_{1}^{2}-\alpha _{3}p_F^{2}a_{1}^{2}+2\alpha
_{3}p_F^{2}a_{3}^{2}\right)A_{4} - \alpha _{1}a_{3}^{2}A_{6}, 
\label{w12} 
\end{eqnarray}%
and 
\begin{eqnarray}
&& w_{21} =\frac{\alpha _{1}}{2}\left(-a_{1}^{2}+3a_{3}^{2}\right)A_{0} \notag\\
\hspace{-0.4cm}&&+\left[L^{2}(\alpha _{1}+p_{F}^{2}\alpha _{3})+\frac{\alpha _{1}}{2}(a_{1}^{2}+a_{3}^{2}) 
-\frac{\alpha_{3}p_F^{2}}{2}(3a_{1}^{2}+a_{3}^{2})\right ]A_{2} \notag \\ 
\hspace{-0.4cm}&&+\frac{\alpha _{3}p_F^{2}}{2}\left(4L^{2}+3a_{1}^{2}+a_{3}^{2}\right)A_{4} -\alpha _{3}p_F^{2}a_{1}a_{3}A_{6},
\label{w21}
\end{eqnarray}%
where the formulas for $A_{0},\ldots,A_{6}$ are provided in Appendix~\ref{integrall}.
Notice that when the cubic Rashba SOC goes to zero ($\alpha_3=0$), Eqs. (\ref{w12w21}, \ref{w12}, \ref{w21}) reproduce the
result derived in Eq.~(\ref{rashbal0}). Furthermore, Eqs.~(\ref{wlc}, \ref{w21}) become zero when $\alpha _{1}=0$, irrespective of $\alpha_3$. As a result we
found that when the linear and cubic RSOC are present, the ISGE is strongly
modified by several new terms in the spin relaxation and the spin generation torques. 

\section{The effects of the linear RSOC and DSOC with the cubic DSOC}

In this section we will evaluate the ISGE in the presence of the linear
Rashba-Dresselhaus SOC combined with the cubic Dresselhaus SOC. To make the
comparison with the experiments easier, we limit ourselves to the
diffusive regime. For the given SOC, the effective SO field ${\bf b}$ is
defined as 
\begin{equation}
\mathbf{b}=p
\begin{bmatrix}
\alpha _{1}\sin \phi +\beta _{1}\cos \phi - p^{2}\beta _{3}\cos 3\phi 
\\ 
-(\alpha _{1}\cos \phi +\beta _{1}\sin \phi +p^{2}\beta _{3}\sin 3\phi)\label{linear_cubic_SOC}%
\end{bmatrix}
\end{equation}%
with $\alpha_{1},\beta _{1}$ and $\beta_{3}$  being the above introduced magnitudes of the linear (Rashba and Dresselhaus) 
and cubic (Dresselhaus) SOC.
According to Eq.~(\ref{termsn1}) and using the form of Eq.~(\ref{linear_spin})
we can show 
\begin{align}
& {}s_{11}=\alpha _{1}\sin 2\phi +(\beta _{1}+\beta _{3}p_F^{2})\cos 2\phi
-2\beta_{3}p_F^{2}\cos 4\phi,   \notag \\
& {}s_{12}=-\alpha_{1}\cos 2\phi +(\beta _{1}-\beta _{3}p_F^{2})\sin 2\phi
-2\beta _{3}p_F^{2}\sin 4\phi,   \notag \\
& {}s_{21}=\alpha_{1}\cos 2\phi +(\beta _{1}-\beta _{3}p_F^{2})\cos 2\phi
+2\beta _{3}p^{2}\sin 4\phi,   \notag \\
& {}s_{22}=\alpha _{1}\sin 2\phi -(\beta _{1}+\beta _{3}p_F^{2})\cos 2\phi
-2\beta _{3}p_F^{2}\cos 4\phi .
\end{align}%
To evaluate the ISGE in the diffusive regime, we need to expand in Eq. (\ref{nth_gk1}) the
denominator~$M_{0}^{-1}$ (with $M_{0}$ presented in Eq.~(\ref{m0})) in terms of the
spin-orbit field. Hence, for all the off-diagonal terms in $M_{0}^{-1}$ we can
neglect $b_{x}^{2}+b_{y}^{2}$ in the denominator with respect to $L$,
whereas for the diagonal terms one must expand the denominator. After this
expansion, the matrix $M_{0}^{-1}$ acquires the form 
\begin{equation}
{M}_{0}^{-1}\approx 
\begin{bmatrix}
1+i\Omega \tau -a^{2}\hat{b}_{y}^{2} & a^{2}\hat{b}_{x}\hat{b}_{y} \\ 
a^{2}\hat{b}_{x}\hat{b}_{y} & 1+i\Omega \tau -a^{2}\hat{b}_{x}^{2} 
\end{bmatrix}.%
\end{equation}%
Now we can derive the Bloch equations (\ref{linear_spin}) 
by inserting the internal magnetic field defined in Eq.~(\ref%
{linear_cubic_SOC}). The spin relaxation rate arises from the left
hand side of Eq.~(\ref{ekeldysh}), which becomes 
\begin{equation}
\hat{\Gamma}=-i\Omega \tau +\hat{\Gamma}_{1}+\hat{\Gamma}_{3}, \label{gamma_t}
\end{equation}%
where the DP spin relaxation for the linear RSOC and DSOC ($\hat{\Gamma}_{1}$)
and the cubic DSOC ($\hat{\Gamma}_{3}$) are given by 
\begin{eqnarray}
\hat{\Gamma}_{1} &=&\frac{a^{2}}{2}\left[ (\alpha _{1}^{2}+\beta
_{1}^{2})\sigma ^{0}+2\alpha _{1}\beta _{1}\sigma ^{x}\right],  \\
\hat{\Gamma}_{3} &=&\frac{a^{2}}{2}\beta _{3}^{2}p_F^{4}\sigma ^{0}.
\end{eqnarray}%
Since the cubic SOC does not produce the spin generation torque
by itself, one can expect that this torque contains the terms originating from the linear coupling and its interplay 
with the cubic one.
Hence, the static limit ($\Omega=0$) of the Bloch equations in Eq.~(\ref%
{linear_spin}) can be rewritten as 
\begin{equation}
(\hat{\Gamma}_{1}+\hat{\Gamma}_{3}){\mathbf{S}}=(\hat{\omega}^{1}+\delta \hat{\omega}%
^{1,3})\hat{\mathbf{E}},
\label{SpinTorque}
\end{equation}%
where the superscript  indices correspond to the linear (1) and cubic parts (3) of the
SOC and to their interplay (1, 3).  By using Eq.~(\ref{m0}) and Eq.~(\ref{termsn1}) the matrices $\hat{\omega%
}^{1}$ and $\delta \hat{\omega}^{1,3}$ appearing in the spin generation
torque in the right hand side of Eq.~(\ref{SpinTorque}) are given by 
\begin{eqnarray}
\hat{\omega}^{1} &=&\frac{S_{0}a^{2}}{2}(\alpha _{1}^{2}-\beta _{1}^{2})%
\begin{bmatrix}
\beta _{1} & -\alpha _{1} \\ 
\alpha _{1} & -\beta _{1}  
\end{bmatrix},
\\
\hat{\omega}^{1,3} &=&\frac{S_{0}a^{2}}{2}%
\begin{bmatrix}
\tilde{\beta}_{1} & -\tilde{\alpha}_{1} \\ 
\tilde{\alpha}_{1} & -\tilde{\beta}_{1}
\end{bmatrix}%
,
\end{eqnarray}%
with 
\begin{eqnarray}
\tilde{\beta}_{1} &=&2p_F^{2}\beta _{3}(2\beta _{1}p_F^{2}\beta _{3}+\alpha
_{1}^{2}), \\
\tilde{\alpha}_{1} &=&\alpha _{1}p_F^{2}\beta _{3}(5p_F^{2}\beta _{3}+2\beta
_{1}).
\end{eqnarray}%
To obtain the spin polarizations, we take the
inverse of the matrix $\hat{\Gamma}$ and multiply it by
the spin generation torque $\hat{\omega}$.  In the zero-frequency limit, $\hat{\Gamma}^{-1}$ is
given by 
\begin{equation}
\hat{\Gamma}^{-1}=\frac{2}{a^{2}}\frac{(\alpha _{1}^{2}+\beta _{1}^{2}+\beta
_{3}^{2}\,p_F^{4})\sigma ^{0}-2\alpha _{1}\beta _{1}\sigma ^{x}}{(\alpha
_{1}^{2}+\beta _{1}^{2}+\beta _{3}^{2}\,p_F^{4})^{2}-4\alpha _{1}^{2}\beta
_{1}^{2}}.
\end{equation}%
Then the spin polarization is defined by 
\begin{align}
\mathbf{S}=S_{0}& \left( \frac{-(\alpha _{1}i\sigma ^{y}+\beta _{1}\sigma
^{z})(\alpha _{1}^{2}-\beta _{1}^{2})^{2}}{(\alpha _{1}^{2}+\beta
_{1}^{2}+\beta _{3}^{2}\,p_F^{4})^{2}-4\alpha _{1}^{2}\beta _{1}^{2}}\right.  
\notag \\
& +\left. \frac{2\beta _{3}p^{2}\,(\xi i\sigma ^{y}+2\,\zeta \sigma ^{z})}{%
(\alpha _{1}^{2}+\beta _{1}^{2}+\beta _{3}^{2}\,p_F^{4})^{2}-4\alpha
_{1}^{2}\beta _{1}^{2}\sigma ^{z}}\right) \hat{\mathbf{E}},  \label{spin_lc}
\end{align}%
where 
\begin{eqnarray}
\xi  &=&\alpha _{1}\Big[\left(-\frac{1}{2}\beta _{3}\,p_F^{2}+2\beta _{1}\right)(\alpha
_{1}^{2}-\beta _{1}^{2})+\beta _{3}\,p_F^{2}(3\beta _{1}^{2}-5\alpha
_{1}^{2})\notag \\
&&-\beta _{3}^{2}\,p_F^{4}(2\beta _{1}+5\beta _{3}\,p_F^{2})\Big], 
\\
\zeta  &=&(\alpha _{1}^{2}-\beta _{1}^{2})
\left(\alpha _{1}^{2}+\frac{1}{4}\beta_{1}\beta _{3}\,p_F^{2}\right)
+\beta _{3}\,p_F^{2}\beta _{1}
(2\beta _{1}^{2}-3\alpha_{1}^{2})  \notag \\
&&+\beta _{3}^{2}\,p_F^{4}(\alpha _{1}^{2}+2\beta _{1}\beta _{3}\,p_F^{2}). 
\end{eqnarray}

In Eq.~(\ref{spin_lc}), the first term corresponds to the linear
Rashba-Dresselhaus SOC, whereas the second one is produced by the presence
of the linear and cubic SOCs and represents their joint action. To compare
our results with Ref.~\cite{paper2_physB}, we begin by considering 
the simple case of $\beta_3=0$. Thus, Eq.~(\ref{spin_lc}) is equivalent to 
Eqs.~(\ref{49}) and (\ref{50}) and the spin polarization is given by
\begin{eqnarray}
\mathbf{S}&=&|e|\tau N_0(\beta_1\sigma^z+\alpha_1i\sigma^y)\mathbf{E} 
\notag \\
&=& \frac{N_0}{2}\mathbf{B}_{\rm int},
\end{eqnarray}
where $\mathbf{B}_{\rm int}$ is the spin-orbit field induced by the
electric current.

\begin{figure}[tbp]
	\centering
	\begin{tabular}{@{}p{\linewidth}@{\qquad}p{\linewidth}@{}}
		\subfigimg[width=\linewidth,pos=ur]{(a)}{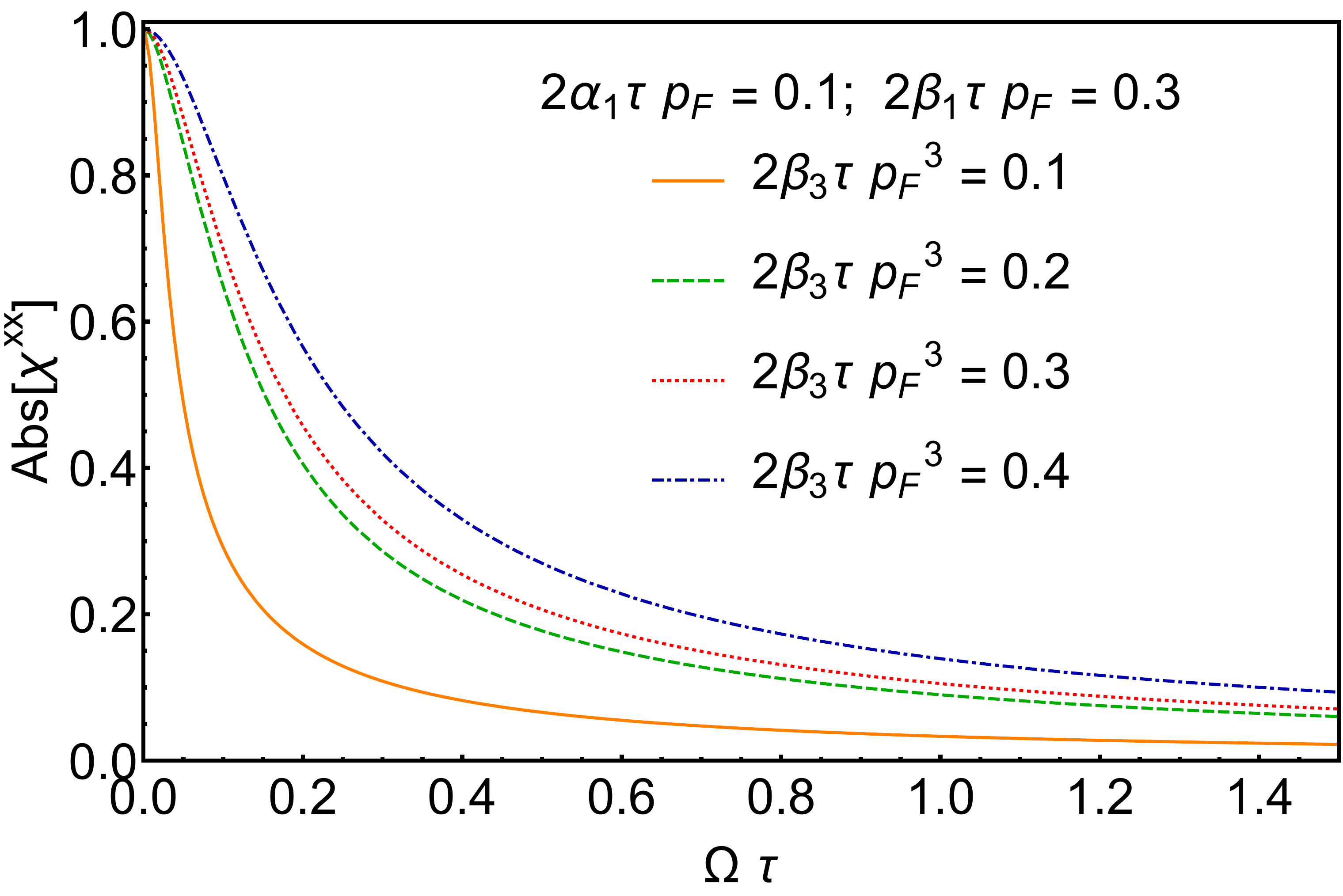} \\
		\subfigimg[width=\linewidth,pos=ur]{(b)}{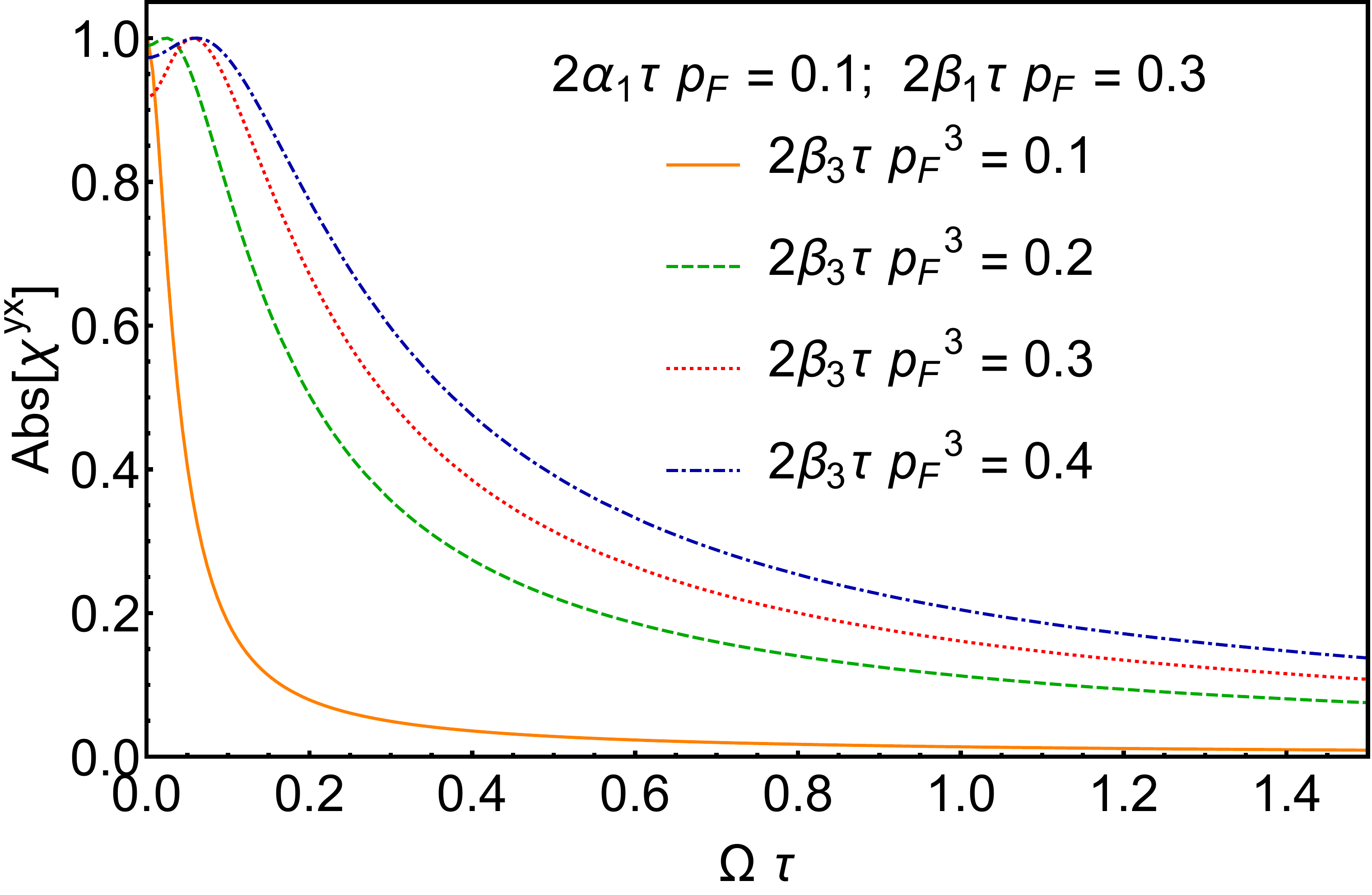} 
	\end{tabular}
		\caption{Absolute value of the normalized SG conductivity as a function of the frequency in diffusive regime. The components (a) $\chi^{xx}$ and (b) $\chi^{yx}$ are induced by the external electric field along \textit{x}-direction. The linear SOC coefficients are fixed: $2\alpha_1\tau p_F=0.1$, $2\beta_1\tau p_F=0.3$. For (a), (b) $2\beta_3\tau p_F^3=0.1$ solid orange, $2\beta_3\tau p_F^3=0.2$ dashed green, $2\beta_3\tau p_F^3=0.3$ dotted red and $2\beta_3\tau p_F^3=0.4$ dot-dashed blue. Results are given in the units of $S_0^{\protect\alpha}$. }
		\label{cubic_linearRD}
\end{figure}

\begin{figure}[tbp]
	\centering
	\begin{tabular}{@{}p{0.7\linewidth}@{\qquad}p{0.7\linewidth}@{}@{\qquad}p{0.7\linewidth}@{}}
		\subfigimg[width=\linewidth,pos=lr]{(a)}{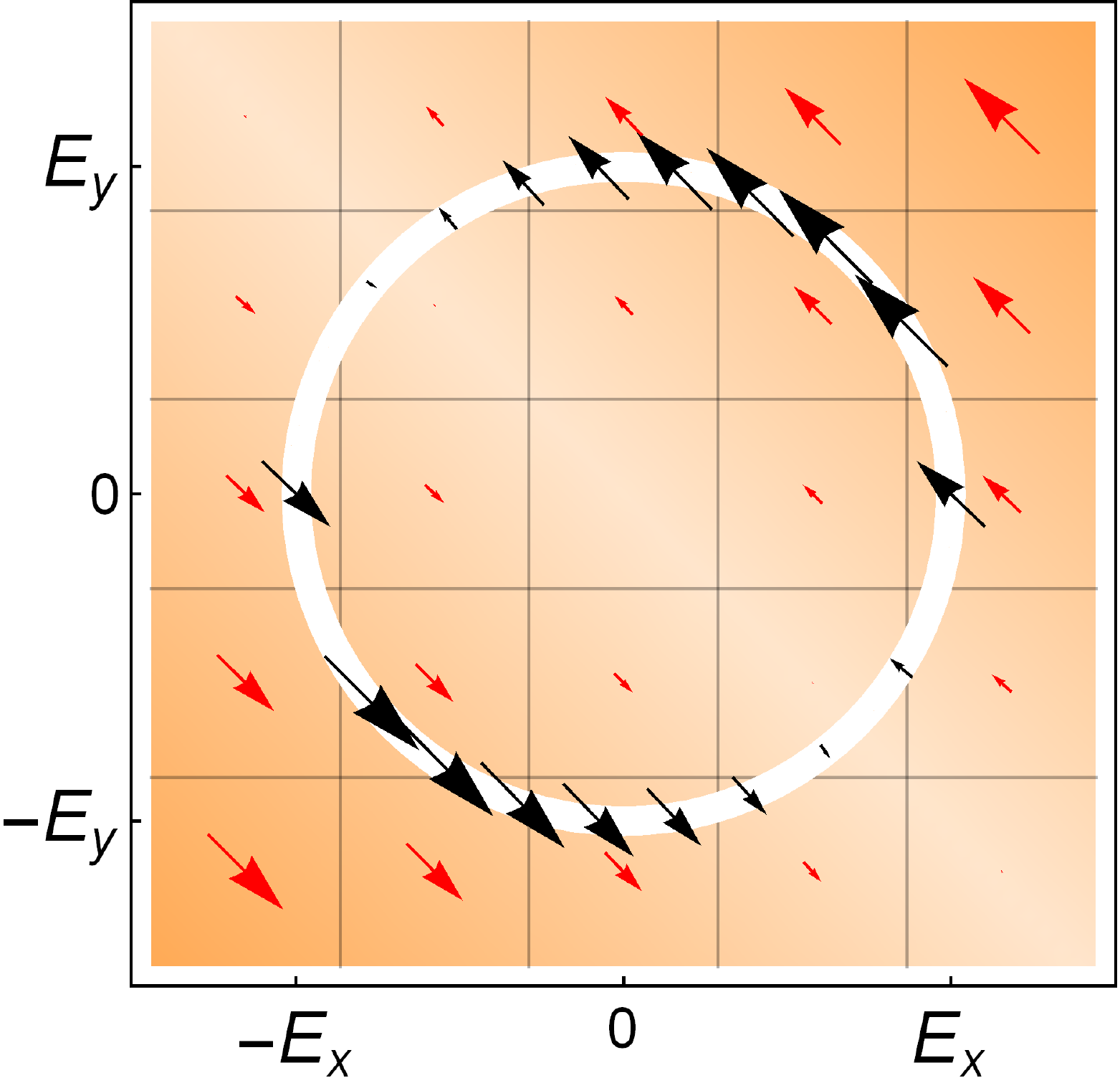} \\
		\subfigimg[width=\linewidth,pos=lr]{(b)}{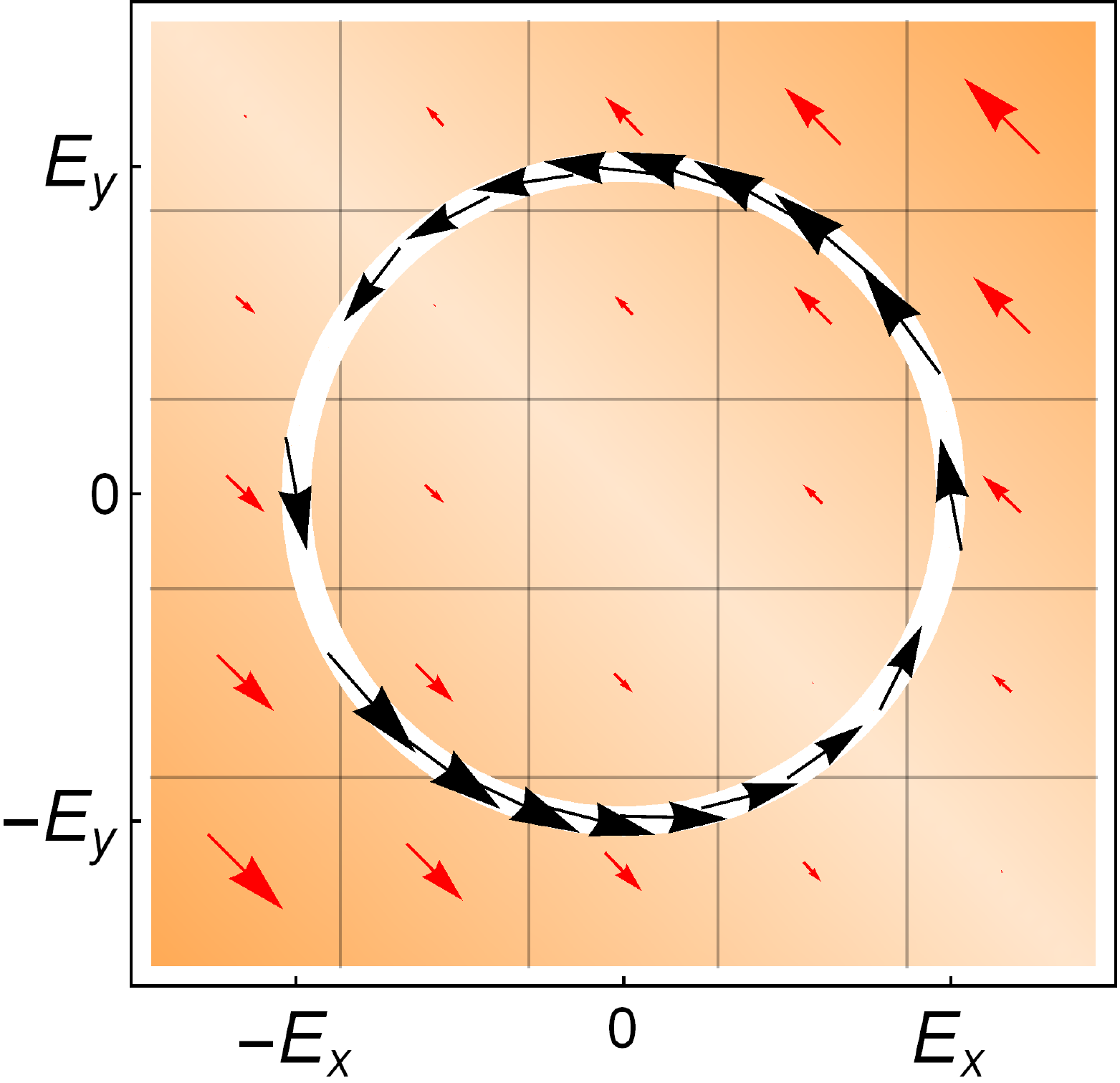} \\
		\subfigimg[width=\linewidth,pos=lr]{(c)}{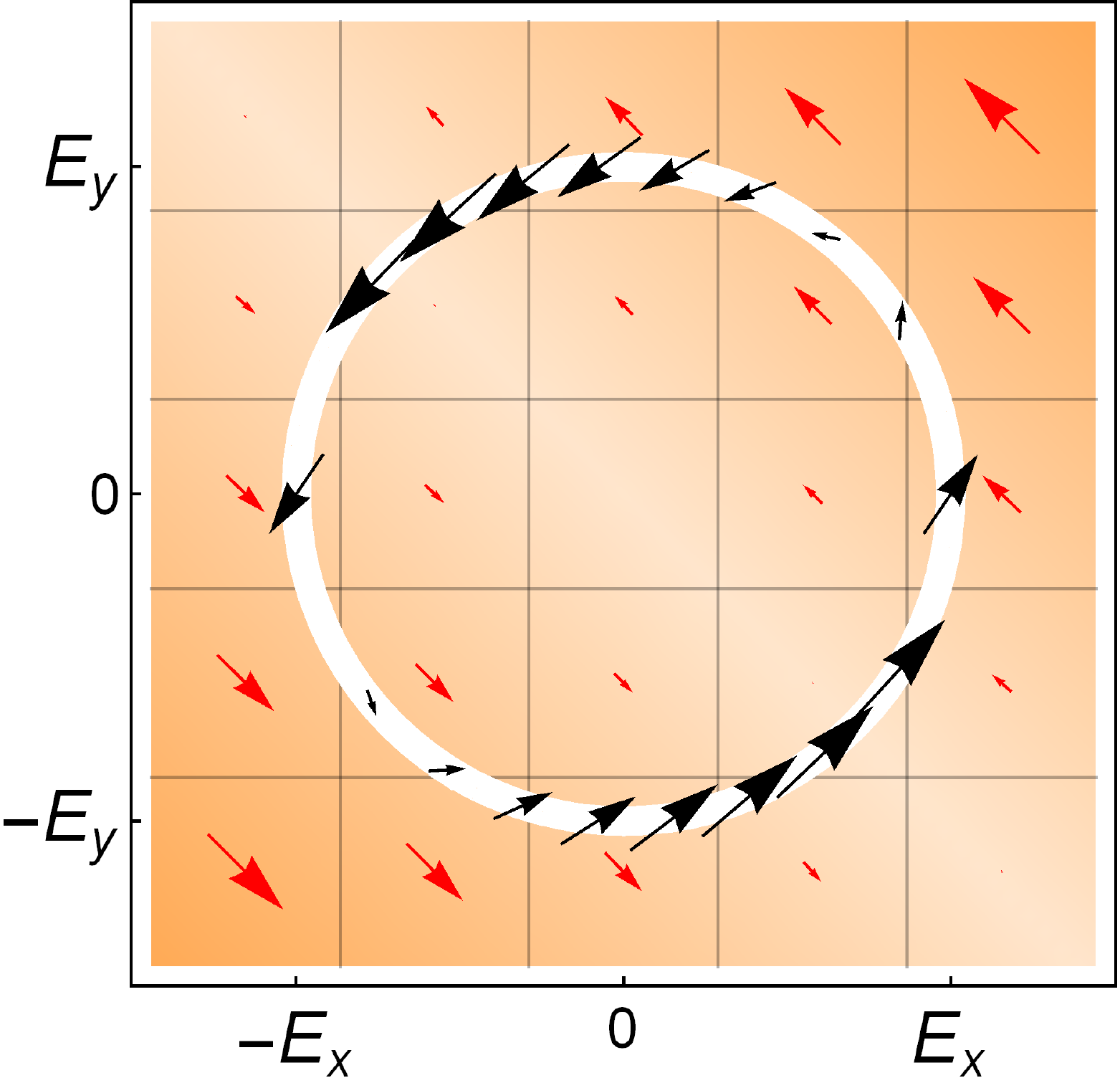}
	\end{tabular}
	\caption{The black arrows within the empty circular sector correspond to the vector plot of the in-plane spin
		polarization ($S^x,S^y$).  The red arrows in the orange background correspond to the direction and the magnitude of the
		magnetic field $B_{\rm int}$, where the
		greatest value is shown by the darkest color of the background.  The results are given in units of $S_0$; $%
		E_{x},E_{y}$ are the components of the electric field $\mathbf{E%
		}$. Linear RSOC and DSOC parameters are $2\alpha_1\tau p_F=0.12$ and $2\beta_1\tau p_F=0.125$.
		(a) Cubic DSOC effect is absent for $2\beta_3\tau p_F^3=0$; (b) cubic DSOC effect, $2\beta_3\tau p_F^3=0.05$, is comparable with the linear RSOC and DSOC effects; (c) cubic DSOC effect, $2\beta_3\tau p_F^3=0.2$, is greater than linear RSOC and DSOC effects.
	}
	\label{vector}
\end{figure}

To analyze the frequency behavior of the CISP, we
consider its real and imaginary components. The imaginary part
originates from $ L=1-i \Omega \tau$ in the matrix $\hat{\Gamma}$, 
whereas the spin generation torque is frequency independent in
the diffusive regime.

The normalized conductivities, $\chi^{xx}$ and $\chi^{yx}$, are shown in
Fig.~{\ref{cubic_linearRD}} as a function of frequency for different
values of the cubic DSOC ($2\beta_3\tau p_F^3$) and fixed values 
$2\alpha_1\tau p_F=0.1$ and $2\beta_1\tau p_F=0.3$. 
We have shown that in the presence of the
linear RSOC-DSOC and cubic DSOC the 
conductivity is the result of the interplay of these two mechanisms.
Besides, the anisotropy of the spin polarization can be controlled by the
strength of the cubic DSOC in addition to the present linear RSOC and DSOC.

To analyze such anisotropy of the ISGE, 
in Fig.~\ref{vector} we present the vector plot of the spin polarization as a function of the 
electric field direction for 
different values of the cubic DSOC. In the two upper diagrams
the ISGE is shown without ($\beta_3=0$, top) 
and with a weak cubic DSOC ($2\beta_3\tau p_F^3=0.1$, middle). One can see that the largest
magnitude of the ISGE occurs for the electric field and the current along the crystallographic direction [1,1] 
(effective linear SOC $\alpha_1+\beta_1$) while the
smallest effect occurs for the field along the  [1,-1] direction (effective linear SOC $\alpha_1-\beta_1$). 
In addition, the bottom plot in Fig.~\ref{vector}  shows that  the increase 
in $\beta_{3}$ ($2\beta_3\tau p_F^3=0.2$)  considerably modifies the anisotropy of the ISGE,
i.e. the strongest polarization is produced now for the spin along the [1,-1] direction and the smallest one 
corresponds to the [1,1] direction.
This picture is  consistent with the experimental result ~\cite{luengo2017,norman2014}.

\section{Conclusions}

In this work, we have studied theoretically the current-induced spin orientation in quantum wells 
by applying the approach based on the quasiclassical Green functions. The theory has been 
developed for the systems where both the linear and the cubic in the electron momentum spin-orbit couplings are present, 
and the corresponding Eilenberger equation have been derived. 
From this equation we obtained the generalized Bloch equations governing the spin 
dynamics of the carriers, permitting us to study a strong spin-orbit coupling  
sufficient to place the spin dynamics beyond the diffusive regime.  
Compared with previous studies, in the case of a sufficiently strong coupling,
we found several new terms arising from the 
interplay of spin-orbit coupling symmetries. For the linear in the momentum coupling, we calculated 
numerically the current-induced spin polarization as a function of frequency of the driving
electric field. We obtained that this polarization can be increased by using the high-frequency 
fields. Since the linear coupling contributes to both the spin generation and spin relaxation torques,
in the static limit the spin polarization always aligns along the internal spin-orbit coupling ``magnetic'' field. 
We noticed that the purely cubic SOC has only effect on the spin relaxation torque, without 
inducing a spin generation torque. When both the linear and the cubic coupling are present, 
the spin generation and spin relaxation are affected differently. As a result, the spin polarization 
is no longer parallel to the spin-orbit field, depending on the relative strength of the linear and cubic couplings. 
This feature agrees with recent experiments~\cite{norman2014,luengo2017}. In general, the approach developed
in this paper can both lead to a better understanding of the spin transport in semiconductors and to a finding of
efficient operational regimes of spintronics devices. 

\section{Acknowledgments}
E.Y.S. acknowledges the support by the Spanish Ministry of Economy, 
Industry, and Competitiveness and the European Regional 
Development Fund FEDER through Grant No. FIS2015-67161-P (MINECO/FEDER), 
and Grupos Consolidados UPV/EHU del Gobierno Vasco (IT-986-16).

\appendix

\section{Explicit matrix form for the Keldysh component of the
quasiclassical Green function}

\label{appmatrixform} 

The matrix forms of the linear operators appearing in
the expressions (\ref{terms}) and (\ref{nth_gk1}) are 
\begin{eqnarray}
M_0= 
\begin{bmatrix}
L & 0 & 0 & 0 \\ 
0 & L & 0 & -2\tau b_0 \hat{b}_y \\ 
0 & 0 & L & 2\tau b_0\hat{b}_x \\ 
0 & 2\tau b_0 \hat{b}_y & -2\tau b_0\hat{b}_x & L%
\end{bmatrix}
,  \label{m0}
\end{eqnarray}
\begin{eqnarray}
N_0+N_1= 
\begin{bmatrix}
1 & -c \hat{b}_x & -c \hat{b}_y & 0 \\ 
-c \hat{b}_x & 1 & 0 & 0 \\ 
-c \hat{b}_y & 0 & 1 & 0 \\ 
0 & 0 & 0 & 1%
\end{bmatrix}%
,\label{n0n1}
\end{eqnarray}
\begin{eqnarray}
S_{\mathbf{E}}= \tilde{E}
\begin{bmatrix}
\hat{\mathbf{E}} \cdot \hat{\mathbf{p}} \\ 
\hat{\mathbf{E}} \cdot \hat{\mathbf{p}} \displaystyle{\frac{N+1}{2} \frac{b_0}{E_F} \hat{b}_x}-
\displaystyle{\frac{\hat{\mathbf{E}}\cdot \partial_{\mathbf{p}}}{v_F}} b_x \\ 
\hat{\mathbf{E}} \cdot \hat{\mathbf{p}} \displaystyle{\frac{N+1}{2} \frac{b_0}{E_F}}\hat{b}_y-
\displaystyle{\frac{\hat{\mathbf{E}}\cdot\partial_{\mathbf{p}}}{v_F}}b_y \\ 
0%
\end{bmatrix}.
\label{termsn1}
\end{eqnarray}

\section{ Integrals over the momentum direction}
\label{integrall}
In this Appendix, we evaluate the integrals used for averaging  over the momentum direction in calculations beyond the 
the diffusive regime. For the combination of linear Rashba and Dresselhaus SOC, one obtains:
\begin{align}
&{}\left\langle \frac{1}{1+\mathcal{C} \sin  2\phi } \right\rangle =  \frac{1}{\sqrt{1-\mathcal{C}^2}},\label{app_lin1}\\
&{}\left\langle \frac{\sin 2\phi }{1+\mathcal{C} \sin 2\phi } \right\rangle = \frac{1}{\mathcal{C}}\left(1- \frac{1}{\sqrt{1-\mathcal{C}^2}}\right),\\
&{}\left\langle \frac{\cos 2\phi }{1+\mathcal{C} \sin 2\phi } \right\rangle = \left\langle \frac{\sin 4\phi }{1+\mathcal{C} \sin 2\phi } \right\rangle=0,\\
&{}\left\langle \frac{\cos 4\phi }{1+\mathcal{C}\sin 2\phi } \right\rangle =
\frac{1}{\sqrt{1-\mathcal{C}^{2}}}+\frac{2}{\mathcal{C}^2}\left(1-\frac{1}{\sqrt{1-\mathcal{C}^2}}\right).\label{app_lin4}
\end{align}

In the presence of both the linear and cubic Rashba SOC, we have the following angular averages: 
	\begin{eqnarray}
	\left\langle \frac{\sin(2n\phi)}{1+\mathcal{D} \cos 2\phi} \right\rangle &= &\left\langle \frac{\sin((2n+1)\phi)}{1+\mathcal{D} \cos 2\phi} \right\rangle=0,\\
	\left\langle \frac{\cos((2n+1)\phi)}{1+\mathcal{D} \cos 2\phi} \right\rangle&=& 0,\quad n=0,1,2,\cdots,\\
    \left\langle \frac{1}{1+\mathcal{D} \cos 2\phi} \right\rangle&= &\frac{- 1}{\sqrt{1-\mathcal{D}^2}}, \\
	\left\langle \frac{\cos 2\phi}{1+\mathcal{D} \cos 2\phi} \right\rangle &= & \frac{1}{\mathcal{D} }\left(1+\frac{1}{\sqrt{1-\mathcal{D}^2}}\right),\\
	\left\langle \frac{\cos 4\phi }{1+\mathcal{D} \cos 2\phi} \right\rangle &= & 
	\frac{1}{\mathcal{D}^2}\left(-2-\frac{-2+\mathcal{D}^2}{\sqrt{1-\mathcal{D}^2}}\right),
	\\
	\left\langle \frac{\cos(6 \phi)}{1+\mathcal{D} \cos 2\phi} \right\rangle &= & 
	\frac{1}{\mathcal{D}^3}\left(4-\mathcal{D}^2+\frac{4-3\mathcal{D}^2}{\sqrt{1-\mathcal{D}^2}}\right).
	\end{eqnarray}

\input{paper_Lin_Cub_Soc_V3_9_04_2018.bbl}

\end{document}

%% file: paper_Lin_Cub_Soc_V3_9_04_2018.bbl
%